\documentstyle[epsf,12pt]{article}
\newcommand{\gtrsim}{\;\raisebox{-0.9ex}
           {$\textstyle\stackrel{\textstyle >}{\sim}$}\;}
\newcommand{\lesssim}{\;\raisebox{-0.9ex}
           {$\textstyle\stackrel{\textstyle<}{\sim}$}\;}
\newcommand{\bold}[1]{\mbox{\boldmath$#1$}}
\newcommand{\wvec}[1]{\stackrel{\longrightarrow}{\rm #1}} 

\newcommand{\missET}{E\llap/_{T}}
\newcommand{\sq}{\tilde{q}}
\newcommand{\glui}{\tilde{g}}
\newcommand{\grav}{\tilde{G}}
\newcommand{\axi}{\tilde{a}}
\newcommand{\neut}{\tilde{\chi}_{1}^{0}}
\newcommand{\chrg}{\tilde{\chi}_{1}^{-}}
\newcommand{\slep}{\tilde{l}^{-}}
\newcommand{\msq}{m_{\tilde{q}}}
\newcommand{\mglui}{m_{\tilde{g}}}
\newcommand{\mgrav}{m_{\tilde{G}}}
\newcommand{\mneut}{m_{\tilde{\chi}_{1}^{0}}}
\newcommand{\Mpl}{M_{\rm P}}
\newcommand{\kgam}{\kappa_{1\gamma}}
\newcommand{\ee}{e^{+}e^{-}}
\newcommand{\rts}{\sqrt{s}}
\newcommand{\Eneut}{E_{\tilde{\chi}_{1}^{0}}}
\newcommand{\Egam}{E_{\gamma}}
\newcommand{\pgam}{\bold{p}_{\gamma}}
\newcommand{\tof}{t_{\rm TOF}}
\newcommand{\thlab}{\theta_{\rm lab}}
\newcommand{\Mgg}{M_{\gamma\tilde{G}}}
\newcommand{\rl}{X_{0}}
\newcommand{\gammas}{\gamma_{*}}
\newcommand{\betas}{\beta_{*}}
\newcommand{\rtF}{\sqrt{F}}
\newcommand{\betaobs}{\beta_{\rm obs}}
\newcommand{\thtun}{\theta_{\rm axis}}
\newcommand{\Enu}{E_{\nu}}
\newcommand{\Emu}{E_{\mu}}
\newcommand{\Lum}{{\cal L}} 
\newcommand{\Fa}{F_{a}}
\begin{document}
\begin{titlepage}
\baselineskip 6mm
\begin{tabbing}
\` UT-ICEPP 97-02 \\
\` {\tt hep-ph/9706382} \\
\` June 1997
\end{tabbing}
\baselineskip 10mm
\begin{center}{\Large\bf 
Hadron colliders as the ``neutralino factory'': \\
Search for a slow decay of the lightest neutralino at the CERN LHC}
\end{center}
\begin{center}{\large 
K.~Maki\footnote{E-mail: maki@icepp.s.u-tokyo.ac.jp.  
Present address:  Hitachi Research Laboratory, Hitachi, Ltd., 
7-1-1 Omika-cho, Hitachi-shi, Ibaraki 319-12, Japan.} 
and S.~Orito\footnote{E-mail: oriton@icepp.s.u-tokyo.ac.jp.}
}\end{center}
\baselineskip 7mm
\begin{center}{\it 
Department of Physics, School of Science, University of Tokyo, 
Tokyo 113, Japan}\end{center}
\vspace{3mm}
\begin{abstract}
\baselineskip 5mm
Prospects are examined for the detection of a slow decay of the lightest 
neutralino (or any other longlived particles) at the CERN LHC and at Very 
Large Hadron Collider (VLHC)\@.  We first point out that such hadron 
colliders will become the ``neutralino factory'' producing $10^{6}$--$10^{9}$ 
neutralinos/yr, if gluinos and/or squarks actually exist below $O(1)$ TeV\@.  
The lightest neutralino ($\neut$), usually assumed to be stable, will be 
unstable if lighter superparticles such as the gravitino ($\grav$) or axino 
($\axi$) exist, or $R$-parity is not conserved.  The decay signal would, 
however, be missed in usual collider experiments, particularly when the decay 
mostly occurs outside the detector.  In order to search for such a slow decay 
of $\neut$, we propose a dedicated experiment where the collision products 
are dumped by a thick shield, which is followed by a long decay tunnel.  The 
decay product of $\neut$ can be detected by a detector located at the end of 
the tunnel.  The slow arrival time and the large off angle (to the direction 
of the interaction point) of the decay product will provide a clear signature 
of slowly decaying $\neut$'s.  One can explore the decay length ($c\tau$) in 
a wide range, {\it i.e.}, 0.2 m to $1\times10^{5}$ km for $\mneut=25$ GeV and 
1 m to 2 km for $\mneut=200$ GeV at the LHC\@.  This corresponds to the range 
of the SUSY breaking scale $\rtF=2\times10^{5}$ to $2\times10^{7}$ GeV in 
case of the $\neut\rightarrow\gamma\grav$ decay predicted in gauge-mediated 
SUSY breaking models.  At VLHC, one can extend the explorable range of 
$\mneut$ up to $\sim1000$ GeV, and that of $\rtF$ up to $\sim1\times10^{8}$ 
GeV\@.  In case of the $\neut\rightarrow\gamma\axi$ decay, the Peccei-Quinn 
symmetry breaking scale $\Fa$ can be explored up to $\sim5\times10^{11}$ 
GeV\@.  The mass of the decaying particle can be determined by using the 
correlation between the energy and the arrival time of the decay product.  
With the setup we propose, one can also search for (i)~other decay modes of 
$\neut$ such as $R$-parity violating one, (ii)~slow decays of any other 
longlived neutral or charged particles, and (iii)~heavy stable charged 
particles.    
\end{abstract}
\end{titlepage}

%
%
\baselineskip 5.9mm
\section{Introduction}\label{sec:intro} 
The search for supersymmetric particles is now an integral part of all 
current, as well as future, experimental programs at high-energy colliders.  
Aside from many attractive features of supersymmetry (SUSY) \cite{G97}, the 
driving force for these searches comes from the recognition that weak-scale 
SUSY, which is introduced to solve the gauge hierarchy problem, requires that 
the SUSY partners of the standard model (SM) particles must be accessible to 
experiments that probe the TeV energy scale.  If this is the case, a large 
number of gluinos and squarks will be produced at future hadron colliders 
such as the CERN LHC (operated at $\rts=14$ TeV with luminosity 
$\Lum=10^{34}$ cm$^{-2}$s$^{-1}$) and Very Large Hadron Collider (VLHC; 
operated at $\rts=100$--200 TeV with $\Lum=10^{34}$--$10^{35}$ 
cm$^{-2}$s$^{-1}$).  Once produced, gluinos and squarks will subsequently 
decay to the lightest neutralino ($\neut$).  This means that such hadron 
colliders will become the ``neutralino factory'', which is capable of 
producing up to $10^{9}$ $\neut$'s per year.    

The $\neut$ is usually assumed to be the lightest supersymmetric particle 
(LSP) and thus stable if $R$-parity is conserved.  It will then escape the 
detector, resulting in the famous missing energy signature for SUSY 
\cite{BCPT95}.  However, the $\neut$ might not be altogether stable:  If 
there exists another superparticle lighter than $\neut$, such as the 
gravitino ($\grav$) or axino ($\axi$), the $\neut$ will decay into, 
{\it e.g.}, $\gamma\grav$ \cite{CFM81} or $\gamma\axi$ \cite{N86}.  
Such a light gravitino naturally exists in gauge-mediated SUSY breaking 
(GMSB) models [5--12] 
as well as in a class of no-scale supergravity (SUGRA) models \cite{EEN84}.  
A light axino can also exist in SUGRA models with the Peccei-Quinn (PQ) 
symmetry \cite{GY92}.  As another possibility, the violation of $R$-parity 
leads to the unstable $\neut$ even if it is the LSP\@.  The $\neut$ will 
then decay into such modes as $q\bar{q}^{\prime}l^{\pm}$, $q\bar{q}\nu$ and 
$\gamma\nu$ \cite{HS84}.  

If the $\neut$ decay takes place inside the detector, the resultant event 
topology would be very different from that in case of the stable $\neut$.  
The experimental signatures have recently been studied for the 
$\neut$ decay into $\gamma\grav$ [16--25] 
and $\gamma\axi$ \cite{HTY97}, motivated by the single 
$ee\gamma\gamma+\missET$ event observed in the CDF experiment at the 
Tevatron \cite{CDF}.  For the CDF event interpreted as the signal of the 
$\neut$ decay, the inferred decay length $c\tau$ is much below 1 m.  However, 
subsequent reports on diphoton $\missET$ distribution observed in the CDF 
\cite{CDF2} and D$\emptyset$ \cite{D0} experiments, as well as the analysis 
of the LEP data at $\rts=161$ GeV \cite{ELN97}, do not give any further 
evidence for the $\neut$ decay into photon(s) with such short $c\tau$.  
Therefore, the possibility is still open for the $\neut$ decay into 
$\gamma\grav$ or $\gamma\axi$ occurring mostly outside the detector.  
Actually, such a slow decay of $\neut$ appears to be favored at least for 
relatively light $\neut$:  Original GMSB models \cite{DN93} prefer relatively 
high SUSY breaking scale, $\rtF\gtrsim10^{7}$ GeV \cite{G97,GMM97}, implying 
$c\tau(\neut\rightarrow\gamma\grav)\gtrsim100$ km for 
$\mneut\lesssim60$ GeV\@.  In case of the $\neut\rightarrow\gamma\axi$ decay, 
the allowed range of the PQ symmetry breaking scale, 
$10^{9}$ GeV$\lesssim\Fa\lesssim10^{12}$ GeV \cite{axion,JKRS96}, leads to 
$c\tau\gtrsim10$ km for $\mneut\lesssim60$ GeV\@.  

If the $\neut$ decay is slow and mostly occurs outside the detector, the 
signature observed in usual collider experiments will be indistinguishable 
from that of the stable $\neut$.  Hence the signal of the $\neut$ 
decay would be missed entirely.  Even if a significant part of $\neut$'s 
produced at hadron colliders decay into photon(s) inside the detector, 
it might be hard to detect the signature with usual detectors, because 
a huge number of particles produced will make it difficult to identify a 
single photon not coming from the interaction point.  In addition, the 
detectors for the approved experiments at the LHC are not designed to 
measure the direction of each photon precisely.  

Therefore, in order to search for a slow decay of $\neut$, we propose a 
dedicated experiment with the interaction point surrounded by a thick 
shield, by which most of the collision products are dumped.  The 
$\neut$ would go through the shield and decay in a long decay tunnel.  
The decay product of $\neut$ can then be detected by a detector located 
at the end of the tunnel.  We show that the slow arrival time and the large 
off angle (to the direction of the interaction point) of the decay product 
will provide an unambiguous signal of slowly decaying $\neut$'s.  We also 
show that, in case of the $\neut$ decay into a photon and a light invisible 
particle such as $\grav$ and $\axi$, the mass of $\neut$ can be determined 
by using the correlation between the energy and the arrival time of the 
decay photon.  Furthermore, by reconstructing the decay kinematics with the 
estimated decay position, one can determine whether the final state is 
two-body or not.  

The remainder of this paper is organized as follows.  We first show in 
Section~\ref{sec:LHC} how the LHC can be the neutralino factory.  In 
Section~\ref{sec:model}, we briefly review the theoretical models of the 
$\neut$ decay into a gravitino or an axino.  Our strategy of the search for 
a slow decay of $\neut$ is then described in Section~\ref{sec:scheme}.  
In Section~\ref{sec:mass}, we illustrate the method of determining 
the mass of $\neut$ after the discovery of its decay.  Section~\ref{sec:VLHC} 
is devoted to the discussion of the $\neut$ decay search at VLHC\@.  We also 
discuss in Section~\ref{sec:other} other searches which are possible with our 
setup, {\it i.e.}, (i)~other decay modes of $\neut$, (ii)~slow decays of any 
other longlived neutral or charged particles, and (iii)~heavy stable charged 
particles.  Finally, we present our conclusions in Section~\ref{sec:conc}.  
A realistic design for the detector is investigated with Monte Carlo 
simulations in Appendix~\ref{sec:dete}.  In Appendix~\ref{sec:munu}, we 
estimate the rates of high-energy prompt neutrinos and muons, which can 
go through the iron shield and thus become the potential background.  

%
%
\section{The LHC as the neutralino factory}\label{sec:LHC}
To simulate the $\neut$ production and decay at the LHC, we first generate 
various SUSY processes in $pp$ collisions at $\rts=14$ TeV using ISAJET 
7.13 \cite{ISAJET} with the CTEQ2L parton distribution functions \cite{CTEQ}.  
In most cases, the dominant products are the gluino ($\glui$) and squark 
($\sq$), which will then decay into $\neut$'s through $\sq\rightarrow\glui q$ 
and $\glui\rightarrow q\bar{q}\neut$ if $\msq>\mglui$, and 
$\glui\rightarrow\sq\bar{q}$ and $\sq\rightarrow q\neut$ if $\msq<\mglui$.  
If the gluino and squark are heavy ($\gtrsim1$ TeV), the production of 
gauginos (charginos and neutralinos) could become dominant.  As an example, 
we take $\msq=1.45\,\mglui$, $\mu=-\mglui$ and $\tan\beta=2$, and assume the 
gaugino mass unification, which is known to roughly hold not only in SUGRA 
models but also in most of GMSB models.

\begin{figure}[htb]
\begin{center}
\epsfbox[60 40 500 500]{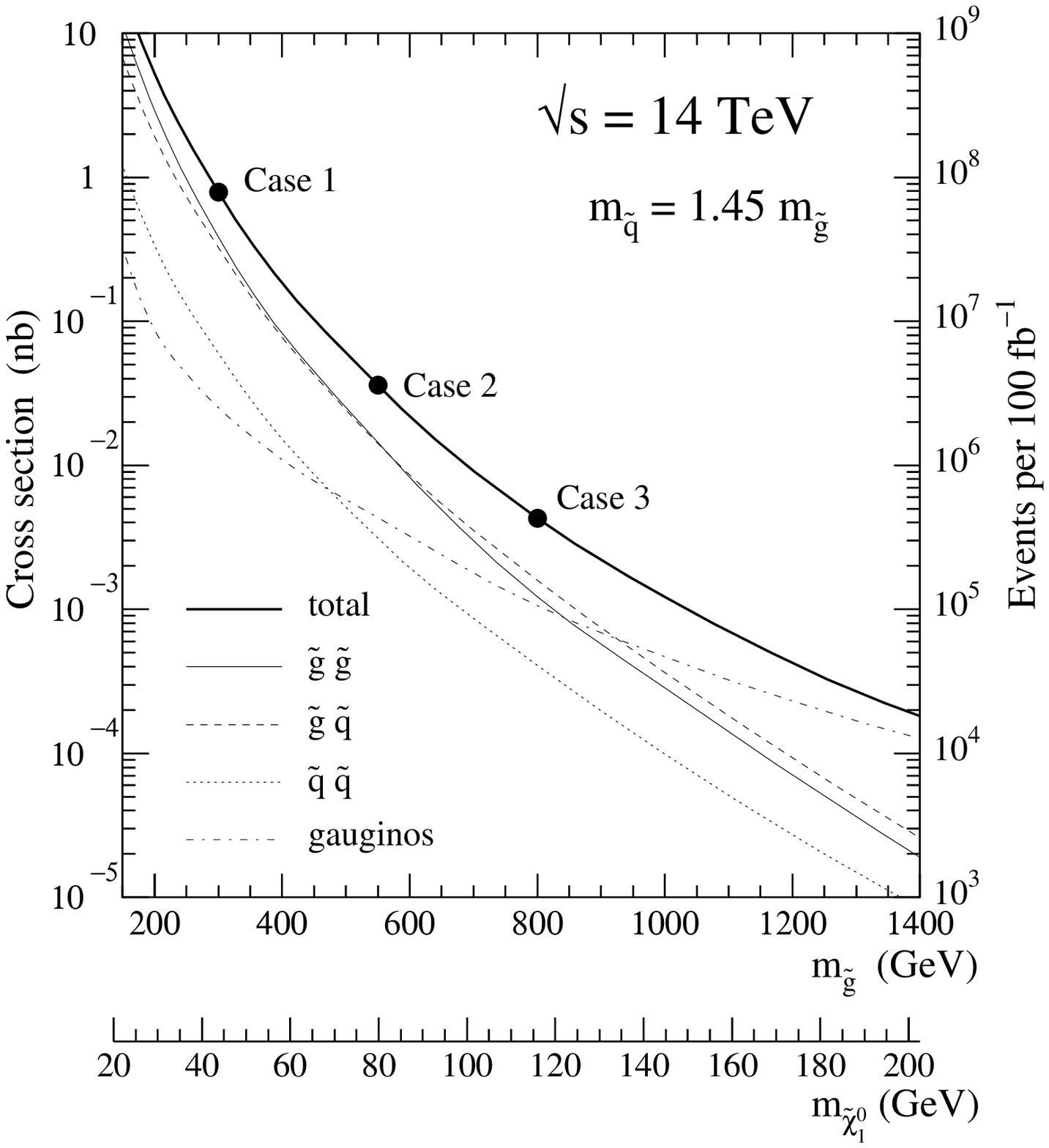}
\caption{
\protect\baselineskip 7mm
Total cross sections for the production of gluinos, squarks and gauginos in 
$pp$ collisions at $\rts=14$ TeV as a function of the gluino mass $\mglui$.  
The corresponding mass of the lightest neutralino $\mneut$ is also shown.  
Also given in the vertical scale is the corresponding number of events per 
100 fb$^{-1}$.}
\label{fig:cross}
\end{center}
\end{figure}

Figure~\ref{fig:cross} shows the total cross sections for the production 
of gluinos, squarks and gauginos, as well as the anticipated numbers of 
events per 100 fb$^{-1}$, at the LHC as a function of $\mglui$.  We note that 
each SUSY event contains two $\neut$'s in the final state.  Also shown 
in this figure are the corresponding values of $\mneut\approx\mglui/7$, which 
results from the gaugino mass unification.  
It is worth noting that the cross sections increase by a factor up to 
$\sim2$ when the next-to-leading 
order QCD corrections are properly included \cite{BHSZ96}.  For the discussion 
below, we select three representative gluino and neutralino masses:  
$\mglui=300$ GeV and 
$\mneut=43$ GeV (Case~1); $\mglui=550$ GeV and $\mneut=78$ GeV (Case~2); 
and $\mglui=800$ GeV and $\mneut=115$ GeV (Case~3).  

\begin{figure}
\begin{center}
\epsfbox[60 30 500 500]{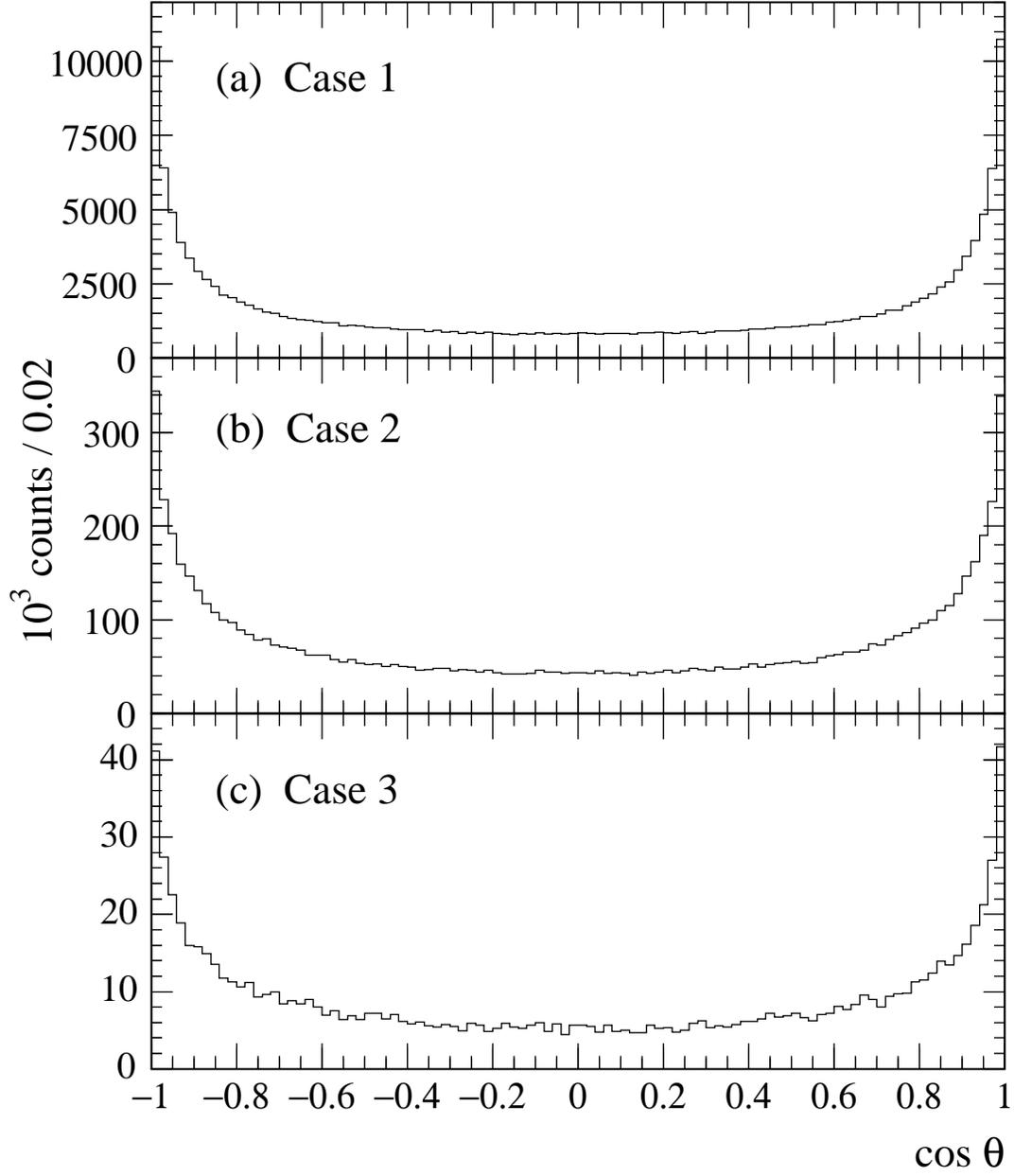}
\caption{
\protect\baselineskip 7mm
Angular distributions of $\neut$'s produced at the LHC with 100 fb$^{-1}$ 
for (a)~Case~1, (b)~Case~2, and (c)~Case~3.}
\label{fig:angle}
\end{center}
\end{figure}

Figure~\ref{fig:angle} shows the angular distributions of $\neut$'s 
produced at the LHC for (a)~Case~1, (b)~Case~2, and (c)~Case~3.  The 
integrated luminosity is taken to be 100 fb$^{-1}$, which corresponds to the 
runtime of $10^{7}$ s with $\Lum=10^{34}$ cm$^{-2}$s$^{-1}$.  The total 
numbers of produced $\neut$'s are $2\times10^{8}$, $8\times10^{6}$ and 
$9\times10^{5}$ for Case~1, 2 and 3, respectively.  It can be seen that 
most of $\neut$'s are produced in the forward direction.  More events can 
thus be detected if one installs a detector at a smaller angle $\theta$ with 
respect to the colliding proton beam.  We here choose $\theta=25^{\circ}$ 
(or $\cos\theta=0.906$) for the central axis of the detector by compromising 
the rate of the signal and that of the potential background (high-energy 
prompt neutrinos and muons) as discussed in Appendix~\ref{sec:munu}.  

Figure~\ref{fig:eneut} shows the energy spectra of produced $\neut$'s 
pointing to the detector which covers $\theta=25^{\circ}\pm10^{\circ}$ 
and an elevation angle of $\pm10^{\circ}$.  We find that, in all three 
cases, the majority of $\neut$'s are produced with relatively low energies, 
{\it i.e.}, the energy spectra have a peak at $\Eneut/\mneut\sim1.5$ to 2.  
This means that most of the produced $\neut$'s are not extremely 
relativistic, having $\beta\equiv v/c\lesssim0.96$.  This makes the time 
measurement of the decay products a powerful tool for the identification 
of slowly decaying $\neut$'s.  

%
%
\section{Models of the lightest neutralino decay}\label{sec:model}
%
%
\subsection{The decay into a gravitino}
Supersymmetry (SUSY) is widely acknowledged to be the best motivated 
extension of the SM\@.  This is not only because it provides a natural 
solution to the gauge hierarchy problem, but also there exist many pieces 
of indirect evidence, such as the correct prediction of $\sin^{2}\theta_{W}$, 
the convergence of the SM gauge couplings at very high energies, and the 
heaviness of the top quark as required by the radiative breaking of the 
electroweak gauge symmetry.  Moreover, SUSY can provide a plausible candidate 
for cosmological dark matter.  These hints, as a whole, appear to constitute 
circumstantial evidence for SUSY\@.   
      
SUSY, however, must be broken because no SUSY partners of ordinary 
particles have been discovered.  The breaking of SUSY is usually 
conceived to occur in a hidden sector at some very high energies, which 
will probably remain unaccessible to direct collider experiments at least 
in the near future.  Thus the investigation into the origin of the SUSY 
breaking is of great importance since it could be the unique window to look 
into the world of such high energies, perhaps up to the Planck scale.  
Two scenarios have been proposed on how the SUSY breaking is communicated 
from the hidden sector to the observable sector:  In one scenario, as in 
supergravity (SUGRA) models \cite{N84}, SUSY breaking is mediated by gravity.  
In this case, the SUSY breaking scale $\rtF$ is necessarily of order 
$10^{11}$ GeV to give the superparticle masses around the weak scale.  The 
other is the gauge-mediated SUSY breaking (GMSB) scenario [5--12], 
where SUSY is broken at a scale as low as $10^{5}$--$10^{7}$ GeV, and the 
gauge interactions act as the messengers of SUSY breaking.  Because of this 
relatively low value of $\rtF$, the mass degeneracy among squarks or sleptons, 
which results from the fact that gauge interactions are flavor blind, is 
hardly broken by the evolution to the weak scale, in contrast to SUGRA 
models.  This ensures sufficient suppression of flavor changing neutral 
current (FCNC) processes.  In addition, GMSB models may also provide a 
solution to the SUSY $CP$ problem \cite{DNS97}.  
  
These two scenarios also lead to distinctive consequences concerning 
cosmological dark matter.  In SUGRA models, the lightest neutralino 
($\neut$) is usually the LSP, and can be the dominant component of dark 
matter.  On the other hand, GMSB models naturally predict that the 
gravitino ($\grav$) comes out to be the LSP\@.  The gravitino with mass 
$\sim0.5$ keV is claimed to be a good candidate for warm dark matter 
\cite{PP82,BMY96}.  We note that it is practically impossible 
to directly detect the gravitino dark matter in laboratories, in contrast 
to the neutralino dark matter, whose direct detection is being pursued in 
many experiments.  

In this way, for finding the future direction of particle physics and 
cosmology, it is of crucial importance to determine which scenario is 
realized in nature.  As will be shown later, our experiment to search for a 
slow decay of $\neut$ can probe a wide range of $c\tau$, which corresponds to 
$\rtF$ of order $10^{5}$ to $10^{8}$ GeV in case of the 
$\neut\rightarrow\gamma\grav$ decay.  This covers much of the $\rtF$ range 
presently considered in GMSB models.  

The gravitino mass $\mgrav$ is related to $\rtF$ (on the condition of the 
vanishing cosmological constant) by 
\begin{equation}
\mgrav=\frac{F}{\sqrt{3}\Mpl}
=2.13\,\left(\frac{\rtF}{3\times10^{6}\,\,{\rm GeV}}\right)^{2}
\;\;\;\;{\rm keV},  
\label{eq:mgrav}
\end{equation}
where $\Mpl=(8\pi G)^{-1/2}=2.44\times10^{18}$ GeV is the reduced Planck 
mass.  This leads to $\mgrav\simeq0.24$ and 24 keV for $\rtF=10^{6}$ and 
$10^{7}$ GeV, respectively.  
Note that the gravitino heavier than $O(1)$ keV overcloses 
the Universe in the standard cosmology.  Such a heavy gravitino, however, 
is still allowed to be the dark matter if the reheating temperature after 
the inflation is low enough \cite{GMM97,MMY93}.  

The longitudinal component of the gravitino, {\it i.e.}, the Goldstino, 
couples to matter with weak (not gravitational) interaction strength 
proportional to $F^{-1}$.  Through this coupling, the $\neut$ will decay 
to a photon and a gravitino with a partial width of \cite{CFM81}
\begin{equation}
\Gamma(\neut\rightarrow\gamma\grav)=\frac{\kgam\mneut^{5}}{16\pi F^{2}},
\label{eq:width_G}
\end{equation}
where $\kgam$ is the square of the photino admixture of the $\neut$.  
If the $\neut$ is a pure bino, $\kgam=\cos^{2}\theta_{W}\simeq0.77$.  
This partial width is most likely to be the dominant one if the $\neut$ is the 
next-to-lightest supersymmetric particle (NLSP) with not too small $\kgam$, 
resulting in the decay length  
\begin{equation}
c\tau\equiv\frac{\hbar c}{\Gamma}
=80\,\kgam^{-1}\left(\frac{m_{\neut}}{100\,\,{\rm GeV}}\right)^{-5}
\left(\frac{\rtF}{3\times10^{6}\,\,{\rm GeV}}\right)^{4}\;\;\;\;{\rm m}.   
\label{eq:ctau_G}
\end{equation}
For $\rtF\gtrsim3\times10^{6}$ GeV and $\mneut\lesssim100$ GeV, this decay 
length is much longer than a typical detector size, leading that most of the 
decays will occur outside the detector.  It has recently been pointed out 
\cite{G97,GMM97} that such large $\rtF$ ($\gtrsim10^{7}$ GeV) is expected in 
original GMSB models \cite{DN93}.  In addition, other GMSB models [9--11] 
tend to predict that SUSY breaking occurs at relatively high ($\gtrsim10^{6}$ 
GeV) energies, although $\rtF=10^{5}$--$10^{6}$ GeV can naturally fit in with 
a very recent model \cite{INTY97}.  Equation~(\ref{eq:ctau_G}) also indicates 
that the measurement of both $\mneut$ and $c\tau$ will give direct 
information on $\rtF$.  

%
%
\subsection{The decay into an axino}
The Peccei-Quinn (PQ) symmetry was introduced to solve the strong $CP$ 
problem \cite{PQ77}.  This symmetry should be explicitly broken, leading 
to a light pseudoscalar, the axion ($a$), whose mass is related to the PQ 
symmetry breaking scale $\Fa$ by  
\begin{equation}
m_{a}=6.2\times10^{-3}\,\left(\frac{\Fa/N}{10^{9}\,\,{\rm GeV}}\right)^{-1}
\;\;\;\;{\rm eV}, 
\end{equation}
where $N$ is the QCD anomaly factor of the PQ symmetry.  The axion is one of 
good candidates for dark matter owing to its feeble interaction with matter.  
The allowed range of $\Fa$ is confined to $10^{9}$--$10^{12}$ GeV 
\cite{axion,JKRS96} from laboratory experiments as well as from astrophysical 
and cosmological considerations.  
  
If both the PQ symmetry and SUSY hold in nature, there should exist the 
fermionic superpartner of the axion, the axino ($\axi$).  The axino can 
remain light even after the SUSY breaking in SUGRA models \cite{GY92}.  
In particular, the axino with mass $\sim O(1)$ keV could serve as warm 
dark matter \cite{RTW91} as is the case for the gravitino.  

If such a light axino exists, the $\neut$ can decay into $\gamma\axi$.  
In case of the $\neut$ being a pure bino, the decay width can be written 
as \cite{HTY97} 
\begin{equation}
\Gamma(\neut\rightarrow\gamma\axi)
=\frac{25\alpha_{\rm em}^{2}}{1152\pi^{3}\cos^{2}\theta_{W}}
\frac{\mneut^{3}}{(\Fa/N)^{2}}, 
\label{eq:width_a}
\end{equation}
which corresponds to the decay length  
\begin{equation}
c\tau=3.6\,\left(\frac{m_{\neut}}{100\,\,{\rm GeV}}\right)^{-3}
\left(\frac{\Fa/N}{10^{9}\,\,{\rm GeV}}\right)^{2}\;\;\;\;{\rm km}. 
\label{eq:ctau_a}
\end{equation}
The lower limit on the PQ symmetry breaking scale, $\Fa\gtrsim10^{9}$ GeV, 
implies $c\tau\gtrsim10$ km for $\mneut\lesssim60$ GeV\@.

%
%
\section{Strategy of the neutralino decay search}\label{sec:scheme}
\subsection{Dedicated experiment}
We propose a completely new type of collider experiment, which is 
schematically shown in Fig.~\ref{fig:scheme}.  Gluinos and squarks (and 
possibly gauginos) are produced in $pp$ collisions at the interaction point 
and promptly decay into $\neut$'s.  The $\neut$'s then enter the decay tunnel 
after going through the shield, by which most of the collision products are 
dumped.  As a simple example, we here consider an iron shield of 10 m 
thickness, and take the tunnel length to be $L=41.6$ m, although a much 
longer tunnel is certainly possible.  The acceptance of the tunnel is taken 
to be $\theta=25^{\circ}\pm10^{\circ}$ and an elevation angle of 
$\pm10^{\circ}$.  We choose this angle $\theta$ by compromising the rate of 
the signal and that of the potential 
background (see Appendix~\ref{sec:munu}).  Two layers of the anti-coincidence 
hodoscopes possibly with track detectors are installed just on the back end 
of the shield for the rejection of punchthrough muons.  It can also be 
used for the trigger of heavy longlived charged particles.  In addition, the 
scintillation counters and/or track detectors can be installed on the wall 
around the decay tunnel to reject potential background events produced by 
cosmic rays.  If the $\neut$ decays to a photon and a light invisible 
particle in the tunnel, the decay photon will reach the end of the tunnel 
and enter a shower detector which has a spherical front face, and initiate 
the electromagnetic shower.  The incident position, angle, and energy of the 
photon, as well as the arrival time relative to the RF clock of the 
accelerator, are measured by the detector.  

Instead of the bulk iron shield, we can use (at extra costs) an 
``active'' $4\pi$-shield, consisting of segmented calorimeters, magnetized 
iron plates and the muon trackers.  This setup will enable us to measure 
the total (and missing) transverse energy of the event and to tag the 
existence of high-energy muons.  

\subsection{Event simulation and the discovery signal}
The decay probability of $\neut$ within a distance of $x$ is given by 
\begin{equation}
P(x)=1-e^{-x/\beta\gamma c\tau},
\label{eq:prob}
\end{equation}
where $\gamma\equiv\Eneut/\mneut$ is the Lorentz factor, and $c\tau$ is the 
decay length.  As a typical case, the probability 
that the $\neut$ with $\gamma=2$ decays in the tunnel 
(10 m $\leq x\leq$ 51.6 m) is calculated to be 20\% for $c\tau=100$ m, 
and 2.4\% for $c\tau=1000$ m.  

\begin{figure}
\begin{center}
\epsfbox[-10 -10 500 300]{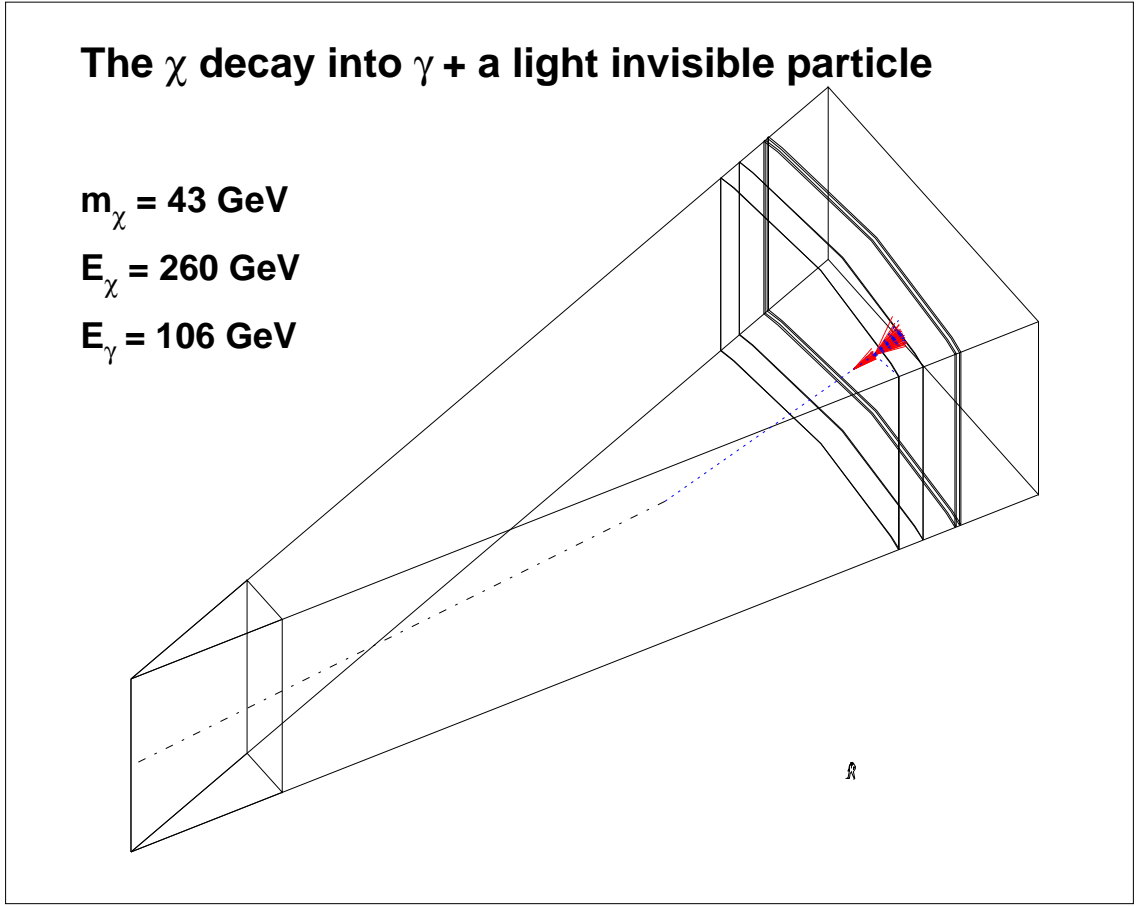}
\caption{
\protect\baselineskip 7mm
An example of the simulated $\neut$ decay into a photon and a light invisible 
particle.  Dash-dotted, dotted and solid lines indicate tracks of the $\neut$, 
$\geq50$ MeV photons and electrons, respectively.}
\label{fig:sample}
\end{center}
\end{figure}

In the following Monte Carlo simulations, we assume that the decay photon 
is emitted isotropically with energy equal to $\mneut/2$ in the rest frame 
of the $\neut$, as is the case for the $\neut\rightarrow\gamma\grav$ decay 
with $\mgrav\ll\mneut$.  The photon 4-momentum ($\Egam$, $\pgam$) in the 
laboratory frame is then calculated by the Lorentz boost.  
We show in Fig.~\ref{fig:sample} a typical generated 
event of the $\neut$ decay into a photon and a light invisible particle.  

\begin{figure}
\begin{center}
\epsfbox[60 30 500 500]{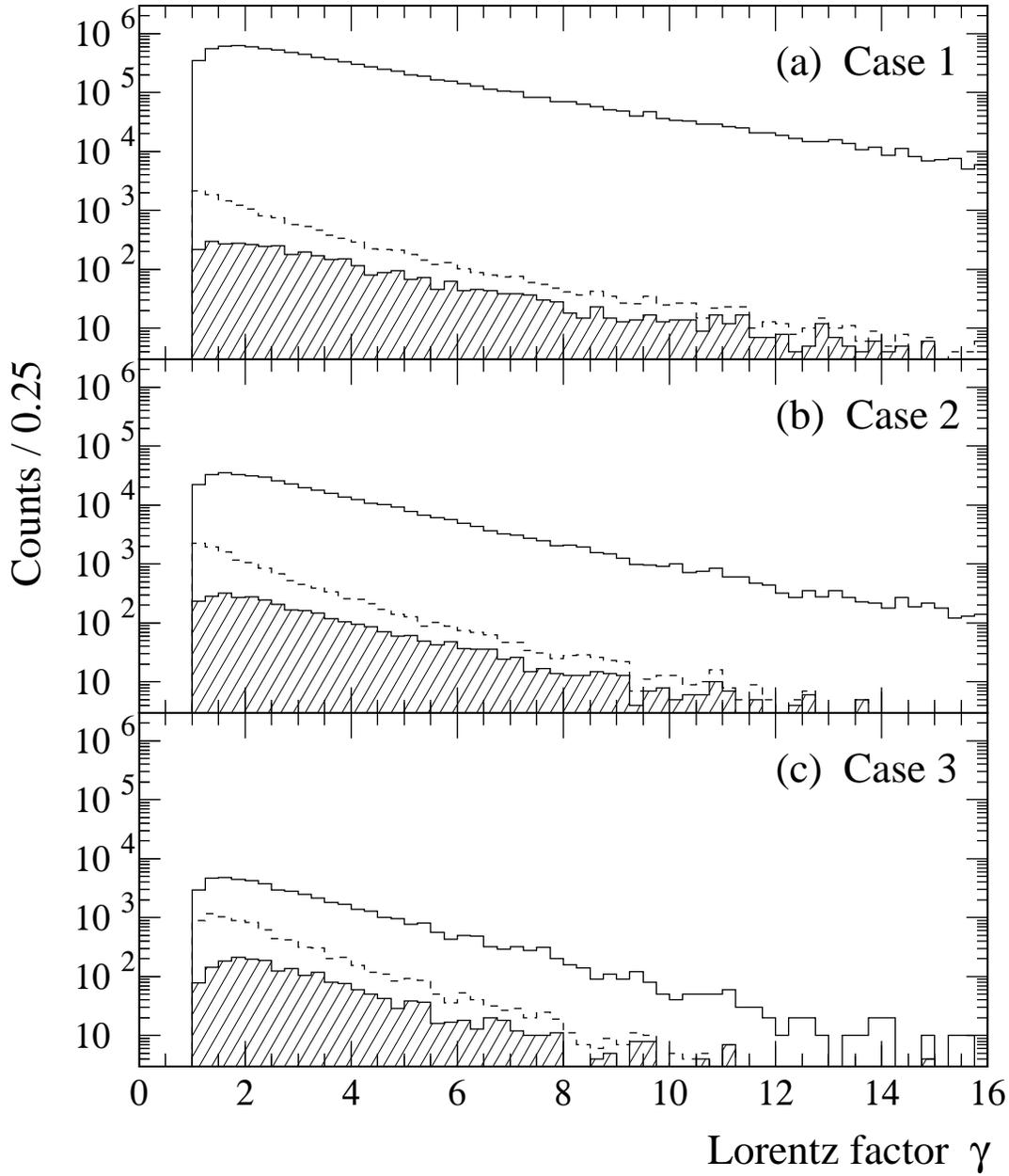}
\caption{
\protect\baselineskip 7mm
Lorentz factor ($\gamma$) distributions of $\neut$'s that enter (solid lines) 
or decay in (dashed lines) the decay tunnel for (a)~Case~1, (b)~Case~2, 
and (c)~Case~3.  Hatched histograms are for $\neut$'s whose decay 
photons enter the shower detector.}
\label{fig:neut}
\end{center}
\end{figure}

\begin{figure}
\begin{center}
\epsfbox[60 30 500 500]{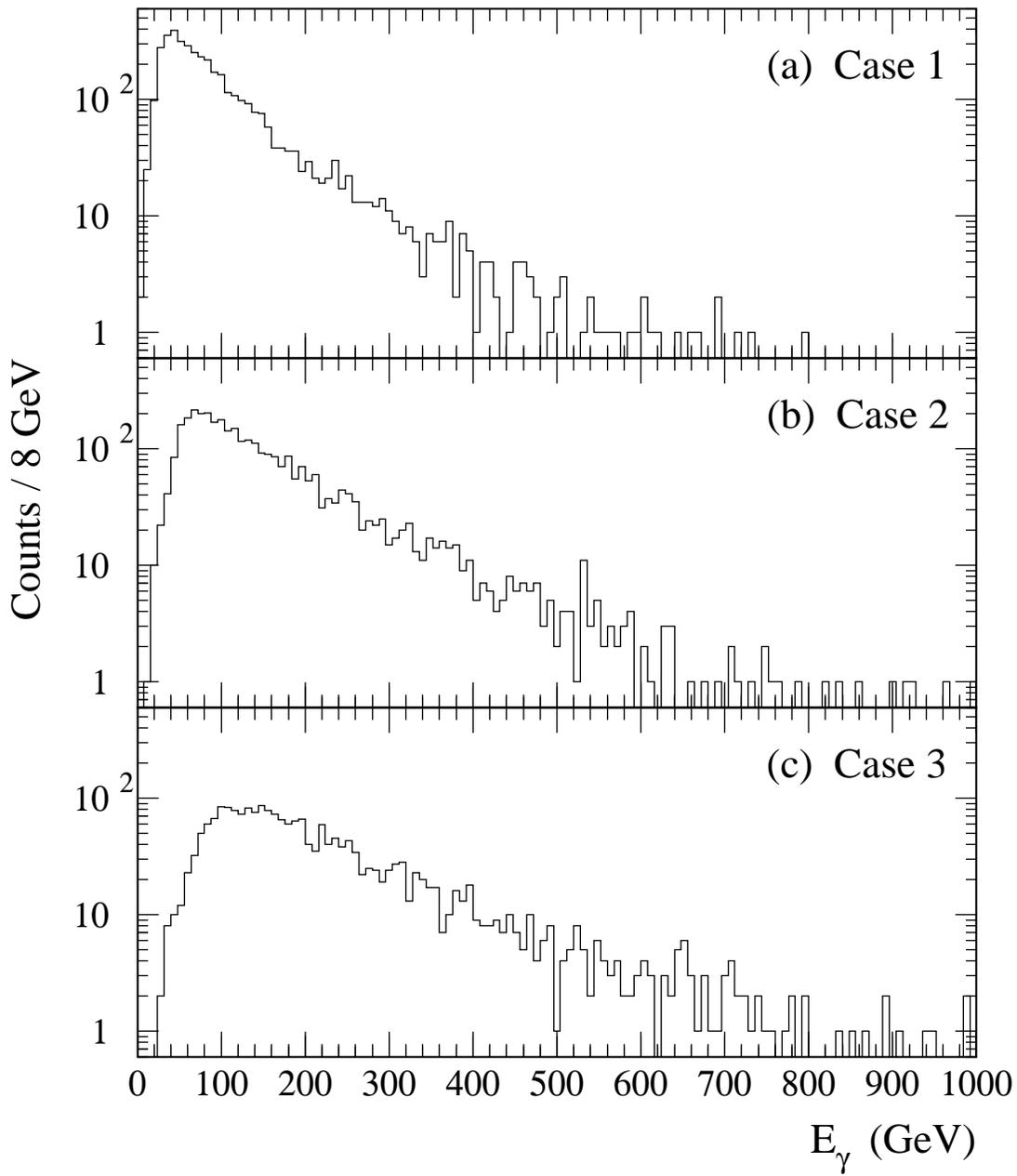}
\caption{
\protect\baselineskip 7mm
The energy spectra of the decay photons that enter the shower detector 
for (a)~Case~1, (b)~Case~2, and (c)~Case~3.}
\label{fig:gamma}
\end{center}
\end{figure}

Figure~\ref{fig:neut} shows the distributions of the Lorentz factor 
$\gamma$ for $\neut$'s entering the decay tunnel for Case~1, 2 and 3 in (a), 
(b) and (c), respectively.  The numbers of $\neut$'s entering the tunnel 
are $9\times10^{6}$, $4\times10^{5}$ and $5\times10^{4}$ 
for Case~1, 2 and 3, respectively, for total integrated luminosity of 100 
fb$^{-1}$, which we assume for all the simulation plots shown in this 
paper.  The distributions of $\gamma$ for 
$\neut$'s decaying in the tunnel are also shown in Fig.~\ref{fig:neut}.  
We here take as an example the decay lengths calculated by 
Eq.~(\ref{eq:ctau_G}) for 
$\rtF=3\times10^{6}$ GeV, {\it i.e.}, $c\tau=7.0$ km, 350 m, and 52 m for 
$\mneut=43$ GeV (Case~1), 78 GeV (Case~2) and 115 GeV (Case~3), respectively.  
The resultant numbers of $\neut$'s decaying in the tunnel are 
$1.4\times10^{4}$, $1.3\times10^{4}$ and $8\times10^{3}$ 
for Case~1, 2 and 3, respectively.  Also shown in this figure by hatched 
histograms are the distributions for $\neut$'s whose decay photons enter the 
shower detector.  The corresponding energy spectra of the decay photons are 
shown in Fig.~\ref{fig:gamma} for (a)~Case~1, (b)~Case~2, and (c)~Case~3.  
The spectra have a peak at $\Egam\sim\mneut$ and extend up to 
$\sim1000$ GeV\@.  
  
The clear separation of the $\neut$ decay signal from the potential 
background, {\it i.e.}, high-energy prompt neutrinos and muons as well as 
the cosmic-ray 
events, can be attained by measuring the arrival time and the off angle 
$\psi$ to the direction of the interaction point (see Fig.~\ref{fig:scheme}).  
Figure~\ref{fig:psitof} shows the scatter plots of $\psi$ vs.\ the arrival 
time for (a)~Case~1, (b)~Case~2, (c)~Case~3, and (d)~background neutrinos 
and muons ($\nu/\mu$).  
The background $\nu/\mu$ events all have $\psi\simeq0^{\circ}$, and arrive 
in narrow bunch structure (here we assume $\sigma_{t}=0.2$ ns).  
On the other hand, most of the decay photons have large $\psi$ (up to 
$\sim30^{\circ}$) and arrive with significant delay (typically 1--20 ns), 
because most of the produced $\neut$'s are not extremely relativistic.  
We show in 
Fig.~\ref{fig:tof} the arrival time distributions.  The evidence for a 
slow decay of the heavy parent particles is clear in the peak position, 
which is displaced by about 2 ns from 
the exact bunch position, as well as in the long tail after the peak.  

The decay length $c\tau$ can in principle be estimated from the distribution 
of the events in the scatter plot of $\psi$ vs.\ the arrival time.  Moreover, 
one could measure $c\tau$ from the variation of the signal count rate 
correlated with the change of $L$, which can be performed by moving the 
detector on rails.  Together with the measured value of $\mneut$ (see 
Section~\ref{sec:mass}), one can determine the strength of the interaction 
which governs the $\neut$ decay.  In particular, 
if one interprets the observed events as the $\neut\rightarrow\gamma\grav$ 
($\neut\rightarrow\gamma\axi$) decay, the SUSY breaking scale $\rtF$ 
(the PQ symmetry breaking scale $\Fa$) can be determined.   

\subsection{Explorable range of the decay length at the LHC}
The large off angle $\psi$ and the slow arrival time are the generic 
features of slowly decaying heavy particles, and will thus always provide 
a clear signature of slowly decaying $\neut$'s.  Based on this prospect, 
we now estimate the explorable range of the decay length $c\tau$.  

\begin{figure}
\begin{center}
\epsfbox[0 30 500 500]{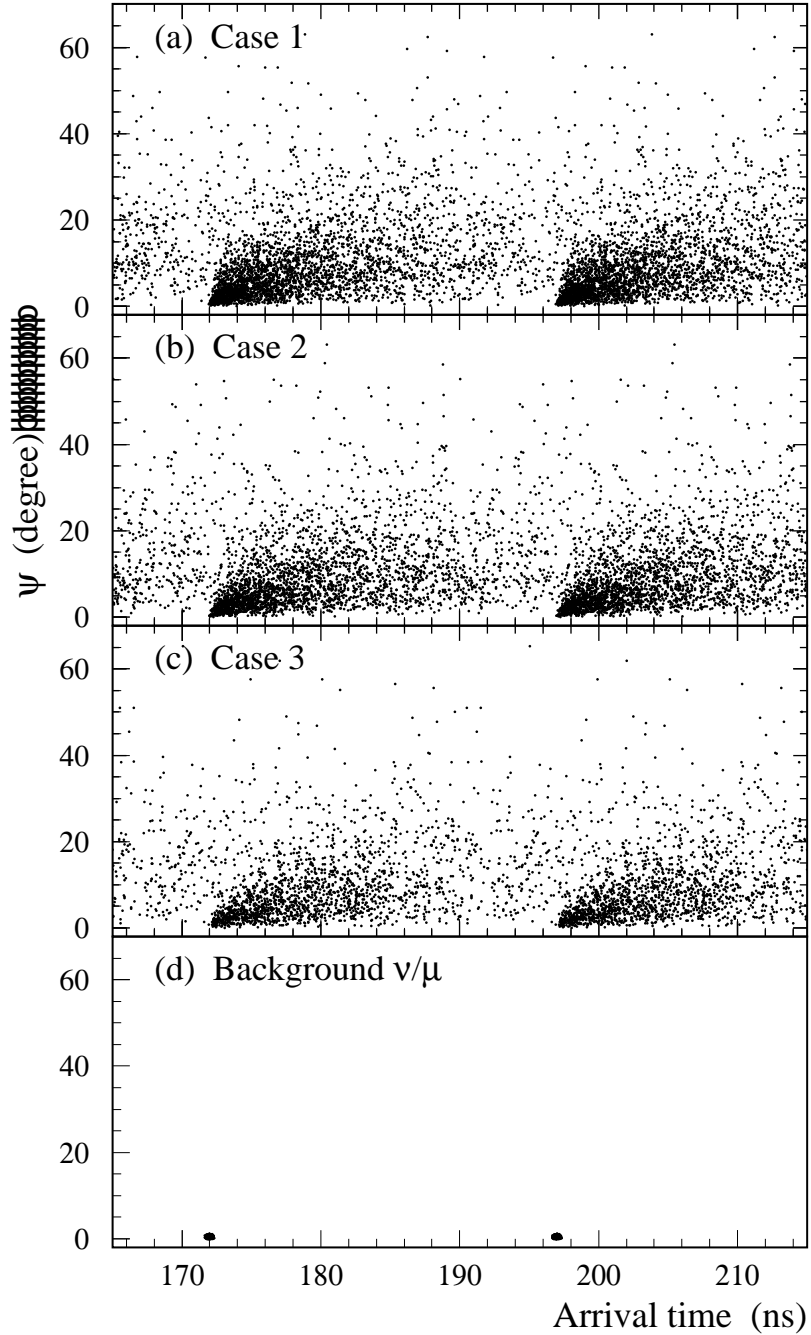}
\caption{
\protect\baselineskip 7mm
Scatter plots ($\psi$ vs.\ the arrival time) of detected events for 
(a)~Case~1, (b)~Case~2, (c)~Case~3, and (d)~background neutrinos 
and muons.  The resolutions of $\psi$ and time measurements are assumed 
to be $\sigma_{\psi}=0.3^{\circ}$ and $\sigma_{t}=0.2$ ns, respectively.}
\label{fig:psitof}
\end{center}
\end{figure}

\begin{figure}
\begin{center}
\epsfbox[0 30 500 500]{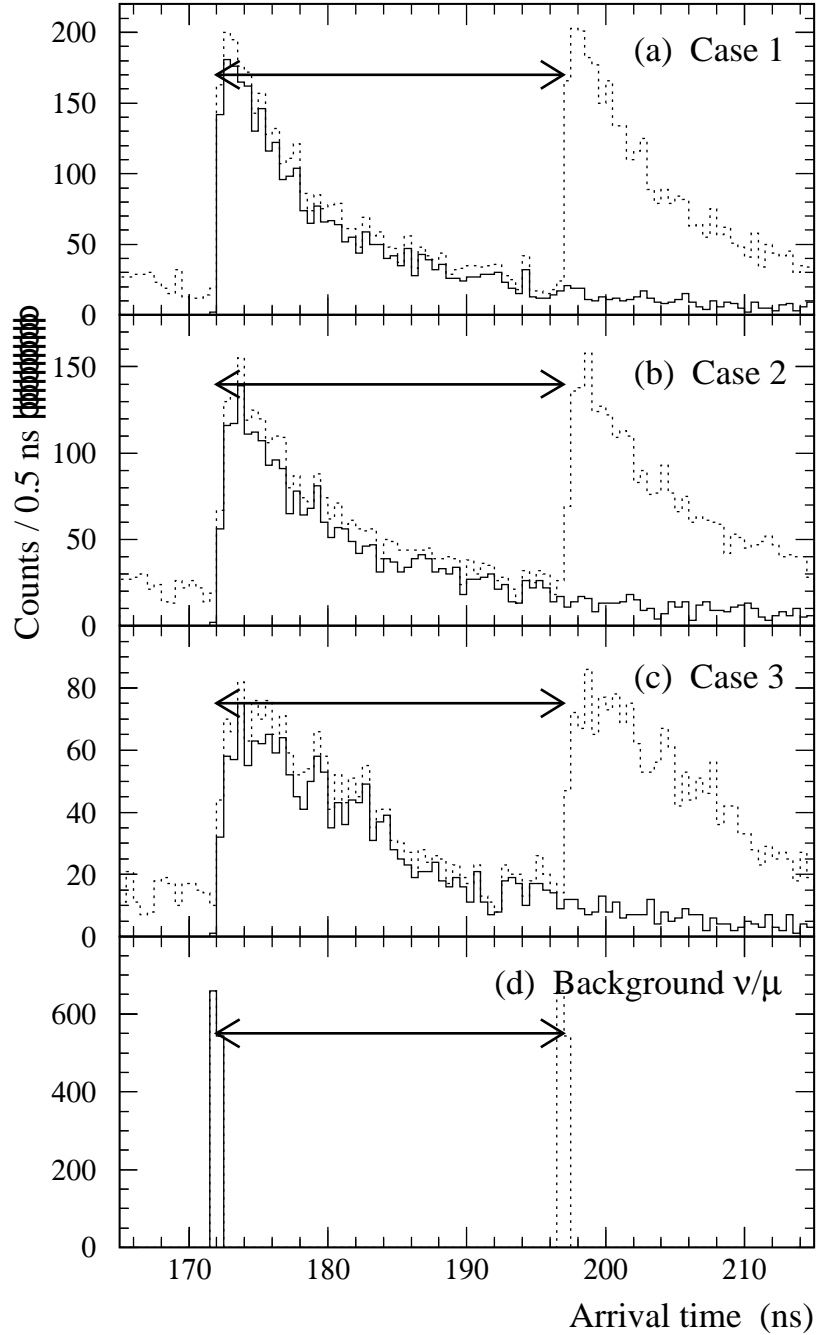}
\caption{
\protect\baselineskip 7mm
Distributions of the arrival time before (solid lines) and after 
(dotted lines) the bunch convolution for (a)~Case~1, (b)~Case~2, 
(c)~Case~3, and (d)~background neutrinos and muons.  
Arrows indicate the bunch spacing of 25 ns.}
\label{fig:tof}
\end{center}
\end{figure}

Figures~\ref{fig:psitof} and \ref{fig:tof} represent typical examples of the 
off angle $\psi$ and the arrival time distributions for $c\tau$ 
much longer than (Cases~1 and 2) or comparable to (Case~3) the tunnel 
length.  For longer $c\tau$, the characteristic shape 
of the distribution will not change, while the number of detected events 
will decrease in proportion to $(c\tau)^{-1}$.  In order to estimate the 
longest explorable decay length, we assume that 10 events with such a 
characteristic distribution are enough to discover the signal.  

For $c\tau$ much shorter than the tunnel length, most of the detected 
events will be due to the $\neut$ decaying immediately after going through 
the shield.  Such events will show the smaller off angle and shorter arrival 
time as compared to the cases shown in Figs.~\ref{fig:psitof} 
and \ref{fig:tof}.  The off angle will nevertheless have finite non-zero 
values (typically of 1 to $10^{\circ}$), and the arrival time will 
definitely be deviated (by 1 to 10 ns) from that of background $\nu/\mu$ 
events, because the $\neut$ flies at least the shield thickness before 
decaying.  In estimating the shortest explorable decay length, we assume 
100 such events are enough to discover the signal.  

\begin{figure}
\begin{center}
\epsfbox[60 40 500 500]{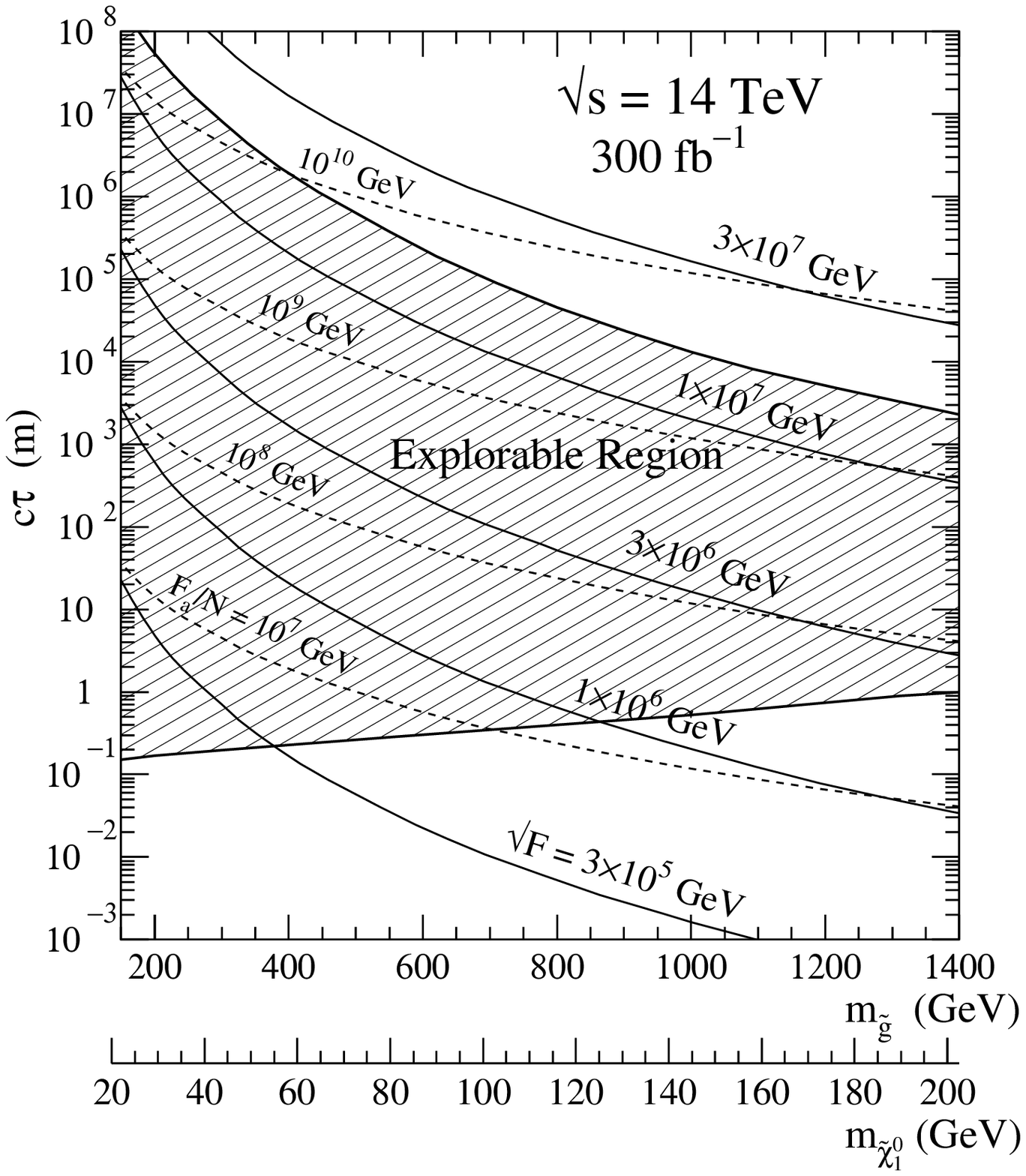}
\caption{
\protect\baselineskip 7mm
Explorable range of $c\tau$ for the $\neut$ decay at the LHC with 300 
fb$^{-1}$.  Also shown are the predicted curves of 
$c\tau(\neut\rightarrow\gamma\grav)$ for 
$\rtF=3\times10^{5}$ to $3\times10^{7}$ GeV (solid lines) and 
$c\tau(\neut\rightarrow\gamma\axi)$ for $\Fa/N=10^{7}$ to $10^{10}$ GeV 
(dashed lines).}
\label{fig:reach}
\end{center}
\end{figure}

Figure~\ref{fig:reach} shows the resultant explorable range of $c\tau$ for 
various $\mneut$ at the LHC with 300 fb$^{-1}$.  
It can be seen that the explorable range of $c\tau$ is wide, 
{\it i.e.}, 0.2 m to $1\times10^{5}$ km for $\mneut=25$ 
GeV, and 1 m to 2 km for $\mneut=200$ GeV\@.  We also show in this figure 
the predicted curves of $c\tau$ for the $\neut$ decay into $\gamma\grav$ and 
$\gamma\axi$ for typical values of $\rtF$ and $\Fa$, respectively.  We find 
that one can explore $\rtF$ in a wide range of $2\times10^{5}$ to 
$2\times10^{7}$ GeV through the $\neut\rightarrow\gamma\grav$ decay.  This 
range includes an interesting case of $\rtF\sim10^{6}$ GeV, where the 
gravitino has mass $\sim0.5$ keV and can be the dominant component of dark 
matter in the standard cosmology.  
In case of the $\neut\rightarrow\gamma\axi$ decay, $\Fa$ can be explored up 
to $\sim10^{10}$ GeV for $\mneut\lesssim60$ GeV\@.

%
%
\section{Determination of the neutralino mass and decay kinematics}
\label{sec:mass}  
After discovering the signal of the $\neut$ decay, one can determine 
the mass of $\neut$ by using the correlation between the energy and the 
arrival time of the decay photon, if enough number of decay events can be 
accumulated.  

\begin{figure}
\begin{center}
\epsfbox[35 270 494 614]{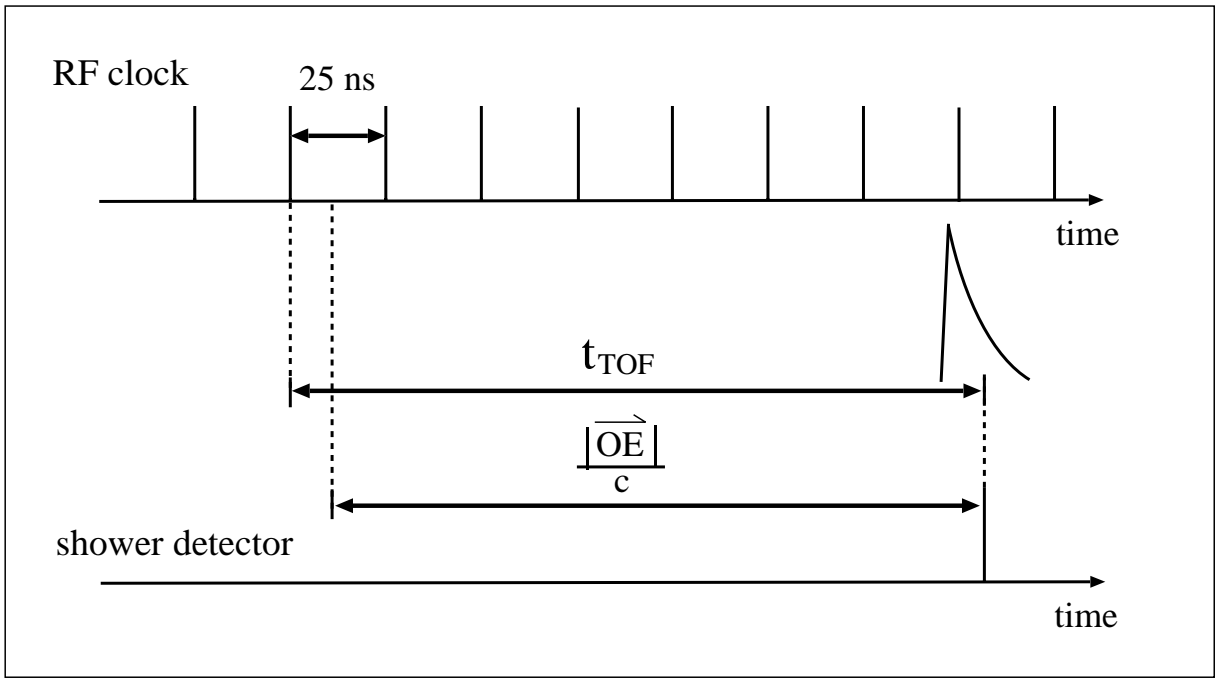}
\caption{
\protect\baselineskip 7mm
Timing chart of the $\neut$ decay event.  The approximate time-of-flight 
(TOF) of $\neut$, $\tof$, is extracted from the time difference between 
the signal of the shower detector and the RF pulse.  Note that 
$|\wvec{OE}|/c$ is the minimum possible TOF on the condition of $\beta\leq1$.}
\label{fig:time}
\end{center}
\end{figure}

\begin{figure}
\begin{center}
\epsfbox[0 30 500 500]{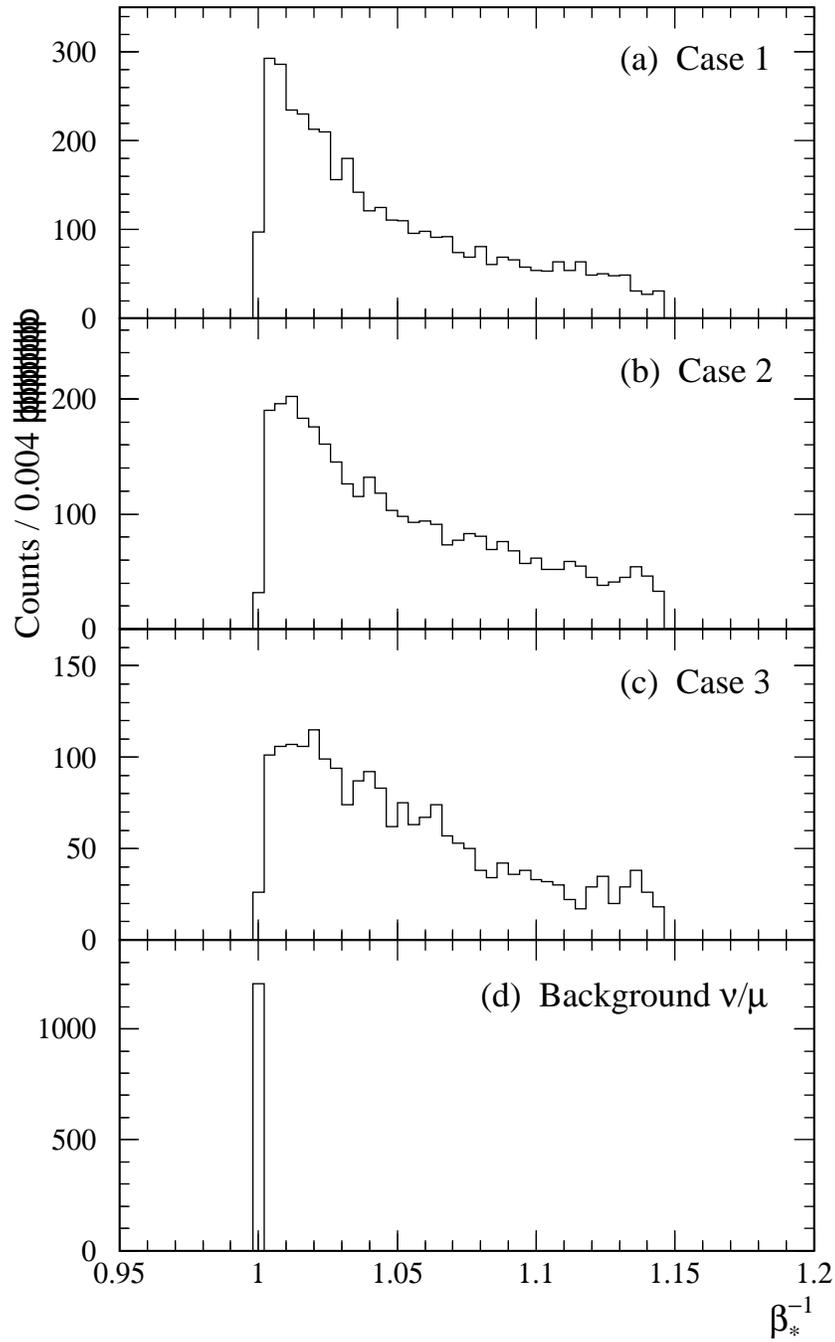}
\caption{
\protect\baselineskip 7mm
Distributions of $\betas^{-1}$ for (a)~Case~1, (b)~Case~2, (c)~Case~3, 
and (d)~background neutrinos and muons.}
\label{fig:beta}
\end{center}
\end{figure}

For this purpose, we first extract the approximate time-of-flight (TOF) 
of $\neut$, $\tof$, from the time difference between the signal of the 
shower detector and the RF clock of the accelerator, as illustrated in 
Fig.~\ref{fig:time}.  In this process, we assume that each $\neut$ comes 
from the closest possible bunch of the proton beam on the condition of 
$\beta\leq1$.  The probability of attributing an event to a wrong starting 
bunch is small, {\it i.e.}, 20--25\% for the three cases described above.  
Using the extracted value of $\tof$, we define the parameter $\betas$ as 
\begin{equation}
\betas\equiv\frac{|\wvec{OE}|}{c\,\tof}, 
\end{equation}
where $|\wvec{OE}|$ is the distance from the interaction point to the front 
face of the shower detector.  The distributions of $\betas^{-1}$ are shown in 
Fig.~\ref{fig:beta} for (a)~Case~1, (b)~Case~2, (c)~Case~3, and 
(d)~background $\nu/\mu$.  There is an upper cutoff at $\betas^{-1}\sim1.15$, 
which results from the modulation of $\tof$ by the bunch spacing of 25 ns.  
The background $\nu/\mu$ events make a sharp peak at 
$\betas^{-1}\sim1$ with no tail in contrast to the $\neut$ decay signal.  
 
\begin{figure}
\begin{center}
\epsfbox[60 30 500 500]{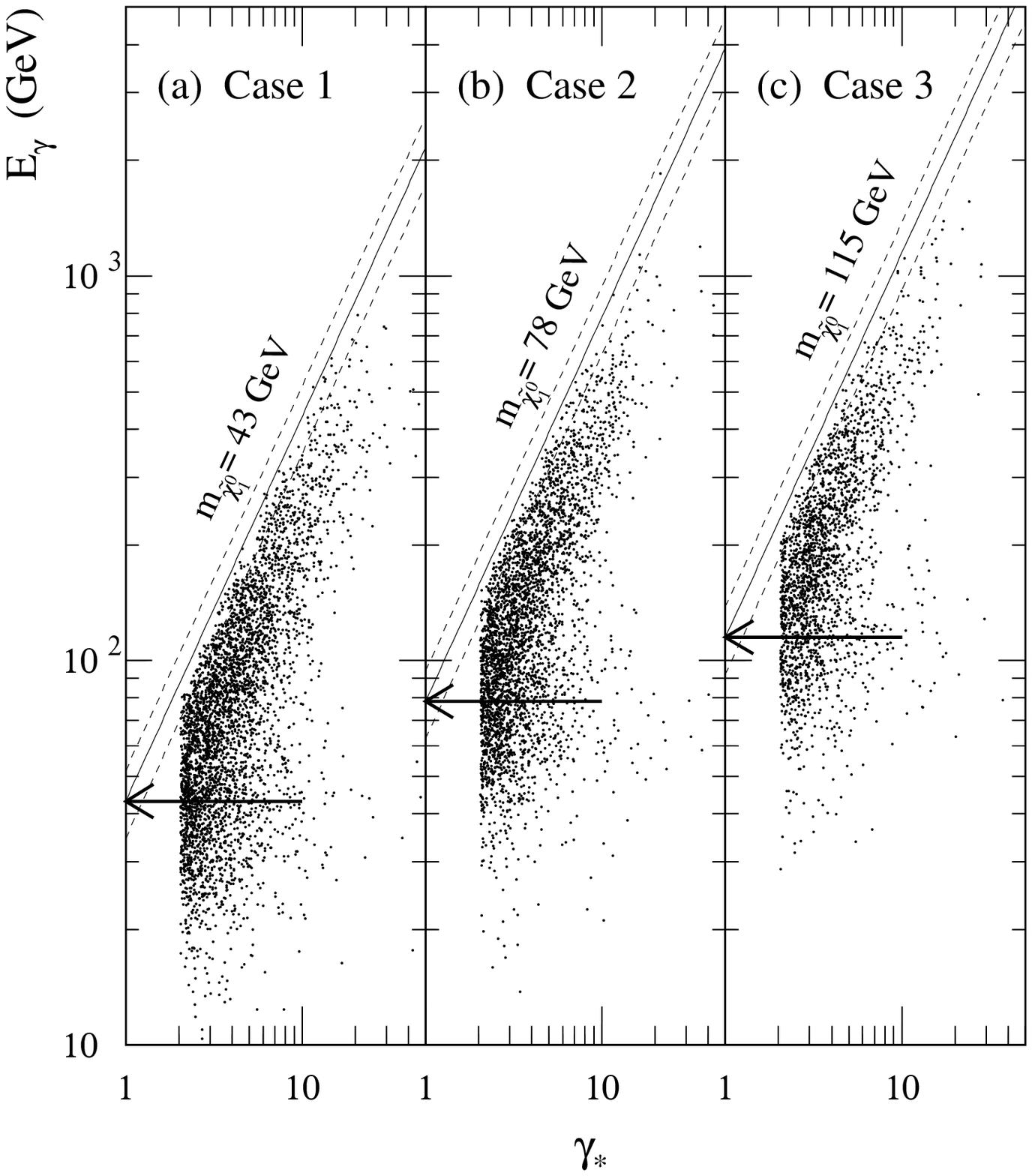}
\caption{
\protect\baselineskip 7mm
Scatter plots of $\Egam$ vs.\ $\gammas$ for (a)~Case~1, (b)~Case~2, 
and (c)~Case~3.  The solid lines correspond to $\Egam=\mneut\gammas$, and 
the dashed lines are for $\mneut$ deviated from the true value by $\pm20$\%.}  
\label{fig:scan}
\end{center}
\end{figure}

\begin{figure}
\begin{center}
\epsfbox[60 30 500 500]{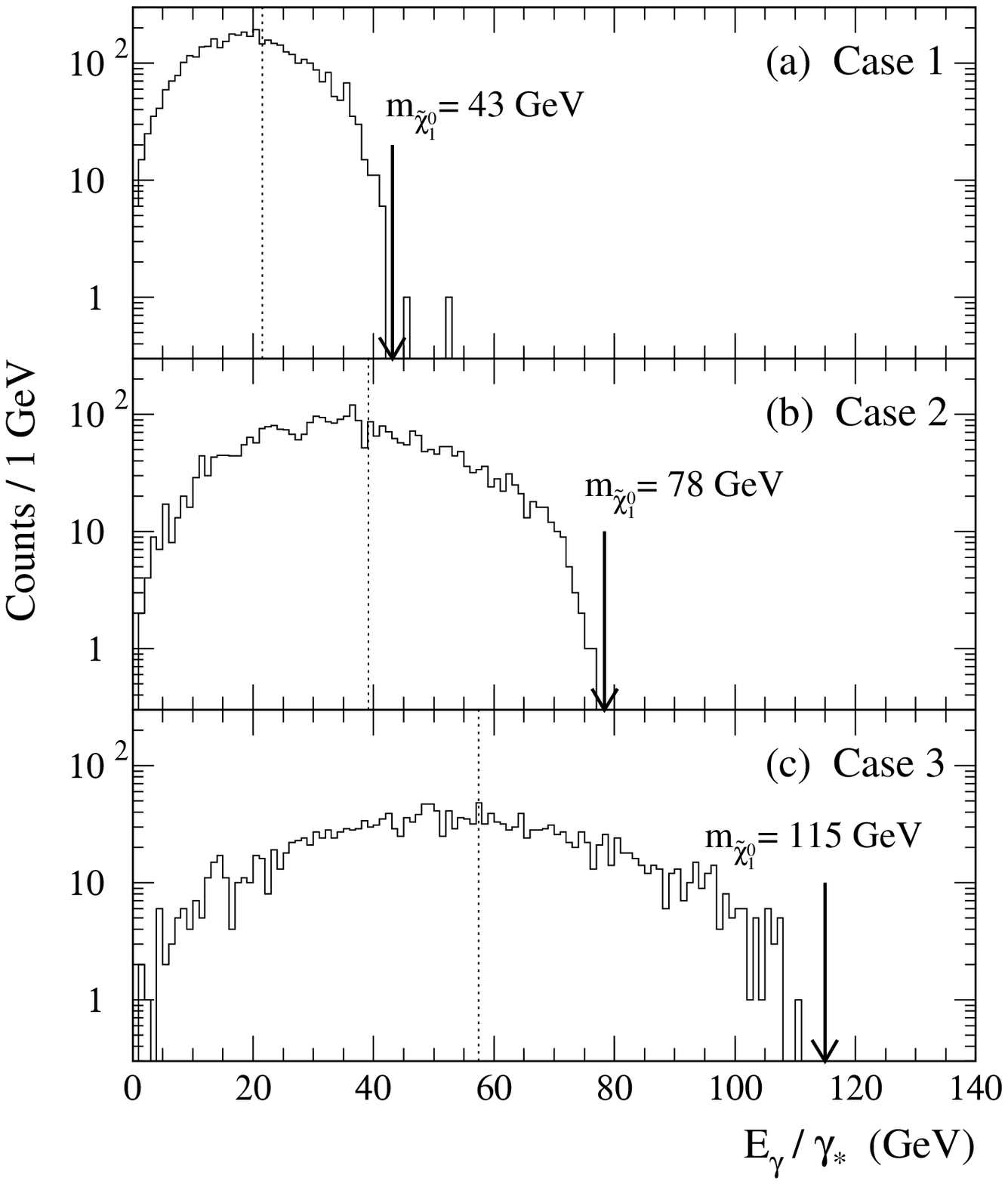}
\caption{
\protect\baselineskip 7mm
Distributions of $\Egam/\gammas$ for (a)~Case~1, (b)~Case~2, 
and (c)~Case~3.  Arrows indicate the position of $\Egam/\gammas=\mneut$.  
The dotted lines show the position of $\Egam/\gammas=\mneut/2$.}
\label{fig:mass}
\end{center}
\end{figure}

Figure~\ref{fig:scan} shows the scatter plots of $\Egam$ vs.\ 
$\gammas\equiv(1-\betas^{2})^{-1/2}$ for (a)~Case~1, (b)~Case~2, and 
(c)~Case~3.  The photon energy resolution is assumed to be 
$\sigma_{E}/E=33\%/\sqrt{E \;({\rm GeV})}$\@.  The lower cutoff of 
$\gammas$ at $\sim2$ corresponds to the above-mentioned cutoff of 
$\betas^{-1}$ at $\sim1.15$.  One can also find an 
edge with constant ratio of $\Egam$ to $\gammas$.  This edge is composed of 
the events in which the decay occurs near the front face of the shower 
detector with the decay photon emitted toward the flight direction of 
the $\neut$.  For such events, $|\wvec{OE}|$ and $\tof$ are nearly the same 
as the actual flight length and TOF of the $\neut$, respectively, and 
the decay photon has the full energy of the $\neut$, {\it i.e.}, 
$\Egam\approx\Eneut$, resulting in the edge corresponding with 
$\Egam/\gammas=\mneut$.  From the position of this edge, the mass of $\neut$ 
can be determined with an accuracy of $\lesssim20$\%.  For this purpose, 
the shower detector should be designed to measure the photon energy in a wide 
range of 10 to 1000 GeV\@.  Figure~\ref{fig:mass} shows the corresponding 
distributions of $\Egam/\gammas$ for Cases~1 to 3.  An upper endpoint 
at $\Egam/\gammas=\mneut$ can be seen in each case.    

\begin{figure}
\begin{center}
\epsfbox[0 30 500 500]{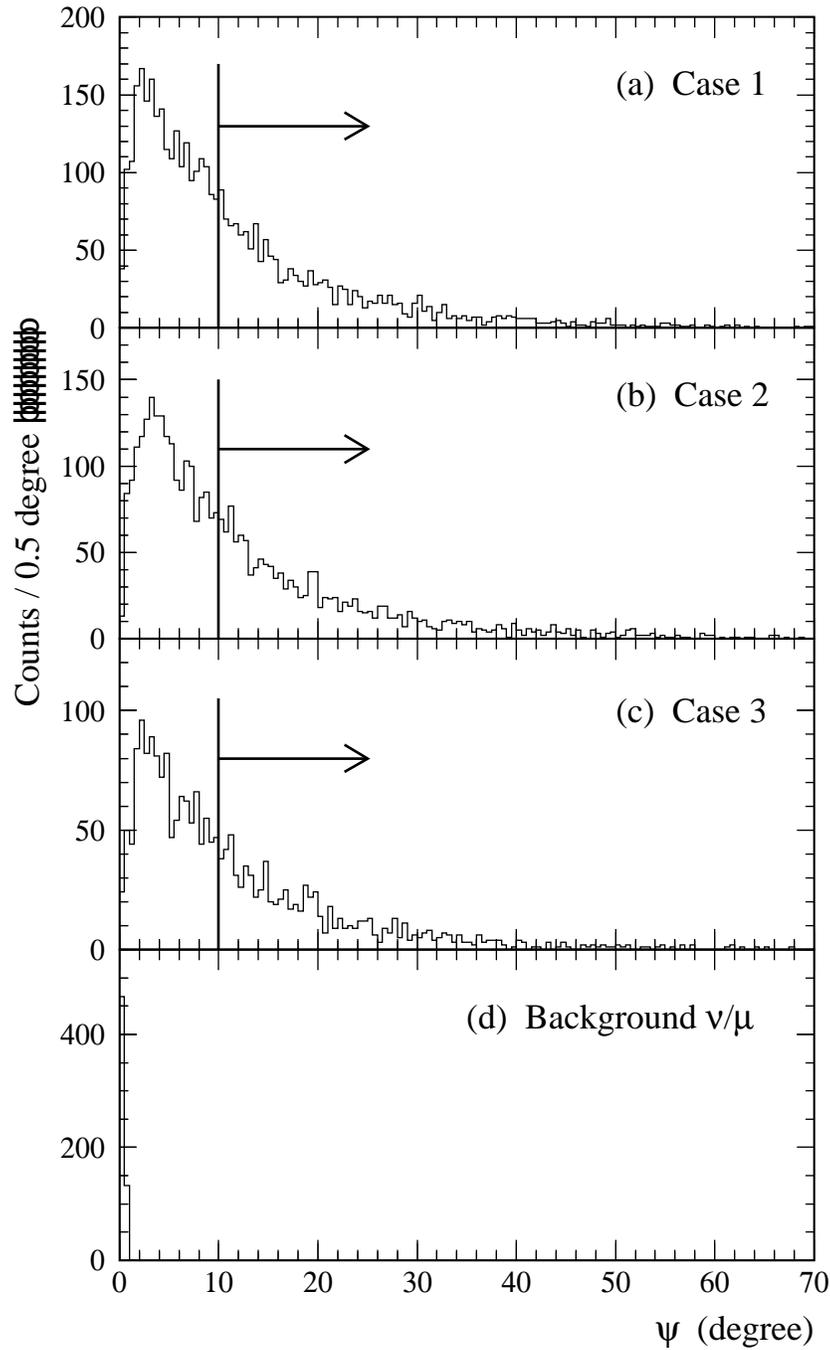}
\caption{
\protect\baselineskip 7mm
Distributions of the off angle $\psi$ for (a)~Case~1, (b)~Case~2, 
(c)~Case~3, and (d)~background neutrinos and muons.  The cut position for 
the kinematical analysis is also shown.}  
\label{fig:psi}
\end{center}
\end{figure}

We now try to determine the nature of the $\neut$ decay, {\it i.e.}, whether 
it is two-body or not.  Although the symmetric peak at 
$\Egam/\gammas=\mneut/2$ in Fig.~\ref{fig:mass} may be the first hint of the 
two-body decay, we can prove the two-body nature by directly reconstructing 
the decay kinematics.  For this purpose, we first select events with 
$\psi\geq10^{\circ}$ based on the $\psi$ distribution shown in 
Fig.~\ref{fig:psi} for (a)~Case~1, (b)~Case~2, (c)~Case~3, and 
(d)~background $\nu/\mu$.  After this $\psi$ cut, only events in which the 
$\neut$ decays relatively near the shower detector can survive.  
The efficiency of this cut is $\sim40\%$ for each of the three cases.  
 
\begin{figure}
\begin{center}
\epsfbox[60 20 500 500]{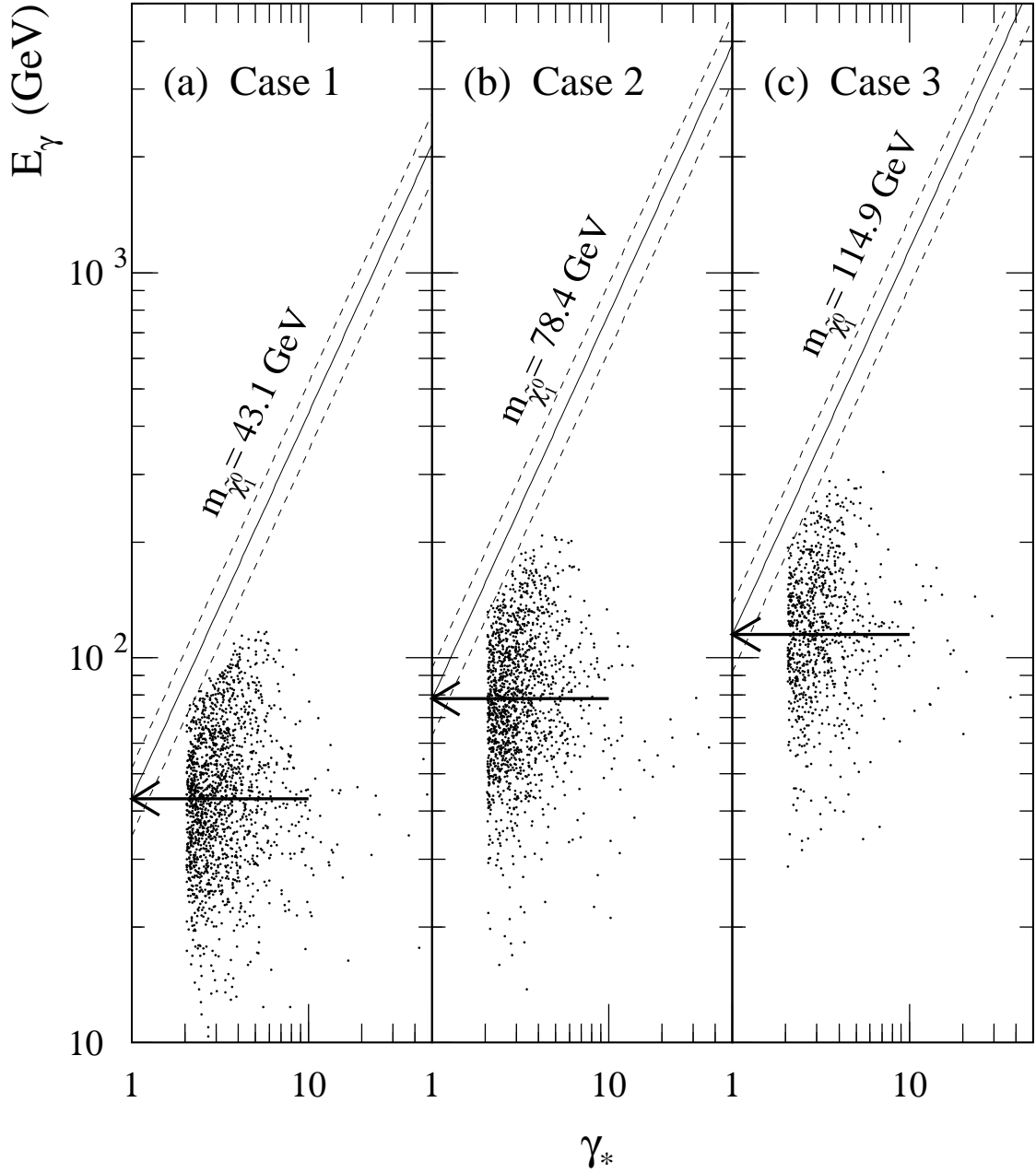}
\caption{
\protect\baselineskip 7mm
Same as Fig.~\protect\ref{fig:scan} after applying the cut of 
$\psi\geq10^{\circ}$.}  
\label{fig:scanc}
\end{center}
\end{figure}

Figure~\ref{fig:scanc} shows the scatter plots of $\Egam$ vs.\ $\gammas$ 
for Cases~1 to 3 after the cut of $\psi\geq10^{\circ}$.  It can be seen 
that only photons with relatively low energies survive.  
Thus the two methods of determining the $\neut$ mass, {\it i.e.}, 
the use of the edge in the scatter plot of $\Egam$ vs.\ $\gammas$, and 
the direct reconstruction of the decay kinematics, are complementary in that 
different events are mainly used in these two methods.  

\begin{figure}
\begin{center}
\epsfbox[19 263 475 555]{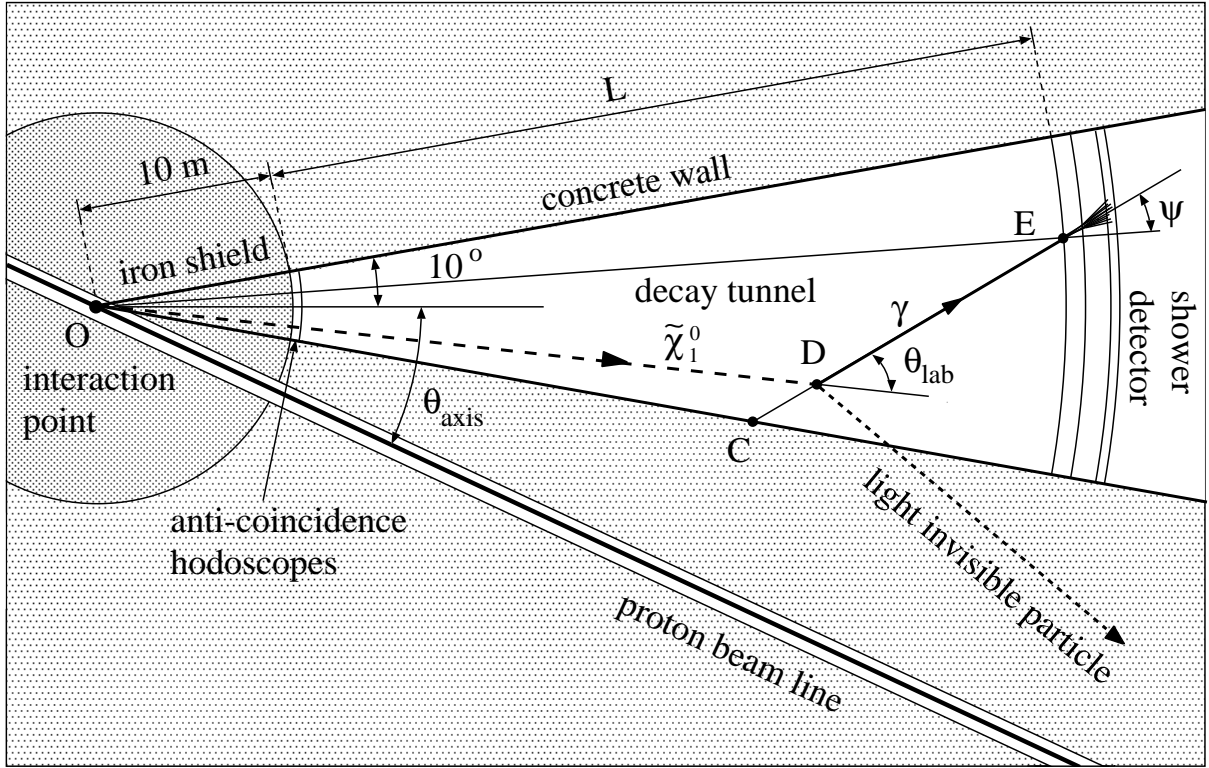}
\caption{
\protect\baselineskip 7mm
Schematic view of the setup.  The lightest neutralino ($\neut$) is produced 
in $pp$ collisions, goes through the iron shield, and decays in the 
tunnel to a photon ($\gamma$) plus a light invisible particle.  The 
decay photon will then enter the shower detector, and be detected.  
An ``active'' $4\pi$-shield consisting of segmented calorimeters, magnetized 
iron plates and the muon trackers can also be used instead of the bulk iron 
shield.}
\label{fig:scheme}
\end{center}
\end{figure}

\begin{figure}
\begin{center}
\epsfbox[60 30 500 500]{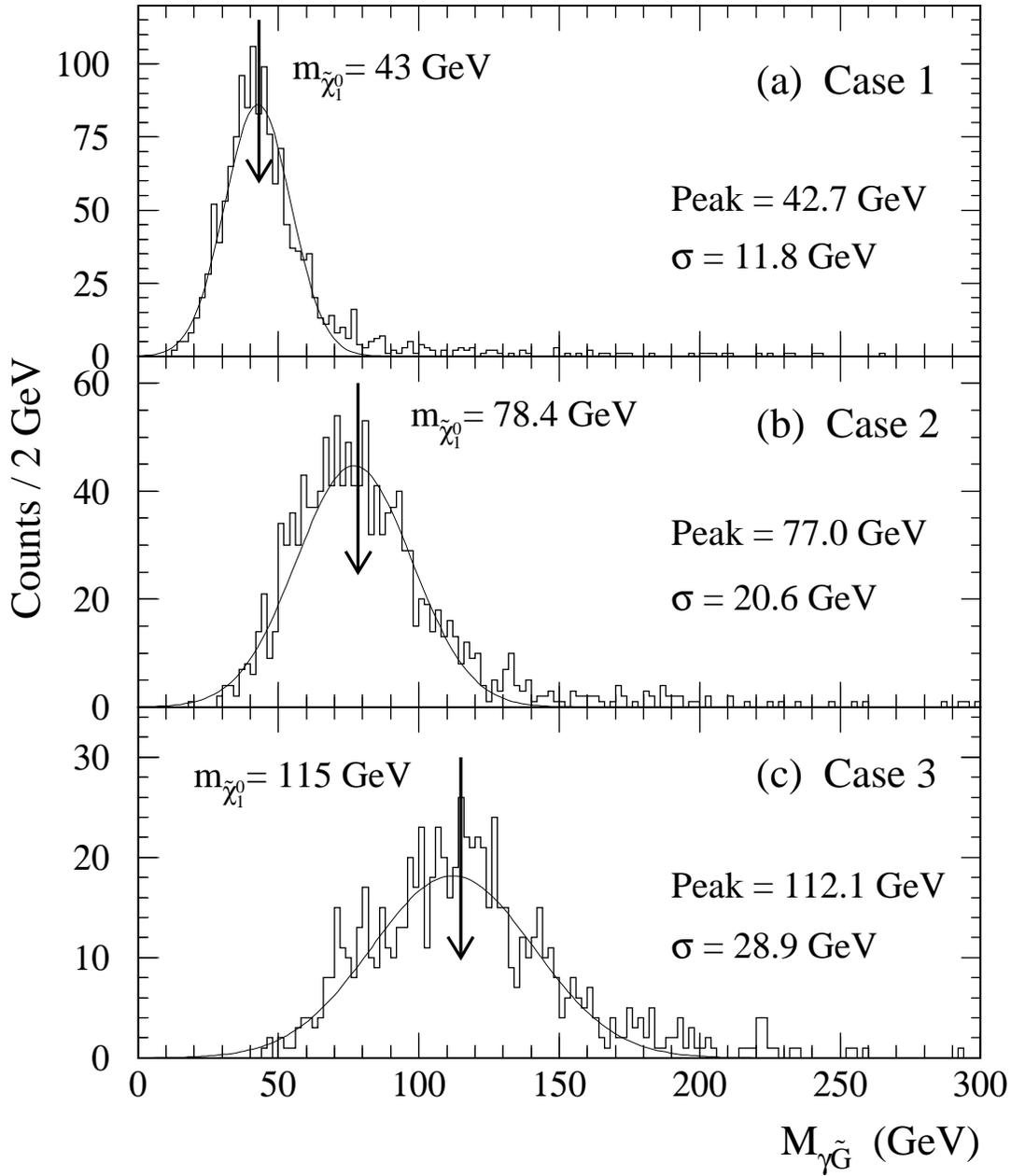}
\caption{
\protect\baselineskip 7mm
Distributions of the reconstructed mass $\Mgg$ for 
(a)~Case~1, (b)~Case~2, and (c)~Case~3.  Solid lines show 
the results of the fit by a gaussian.  Arrows indicate the position of 
$\Mgg=\mneut$.}
\label{fig:recon}
\end{center}
\end{figure}

We then simply assume that the decay position D is the middle point 
between C and E, both of which are defined in Fig.~\ref{fig:scheme}.  
Using this assumption, the velocity of the $\neut$ can be approximately 
expressed by 
\begin{equation}
\betaobs=\frac{|\wvec{OD}|}{c\,\tof-|\wvec{DE}|},
\end{equation}
where $|\wvec{OD}|$ and $|\wvec{DE}|$ represent the flight path of the 
$\neut$ and that of the photon, respectively.  One can then calculate the 
reconstructed mass as 
\begin{equation}
\Mgg\equiv\Egam(1-\betaobs\cos\thlab)(1-\betaobs^{2})^{-1/2},
\label{eq:Mgg} 
\end{equation}
where $\thlab$ is the angle between the momentum of the $\neut$ and that 
of the photon (see Fig.~\ref{fig:scheme}).  Figure~\ref{fig:recon} shows 
the distributions of $\Mgg$ for (a)~Case~1, (b)~Case~2, and (c)~Case~3.  
It can be seen that the $\neut$ mass can be determined with an accuracy of 
$\sim30$\% for each case.  If one finds a peak in the distribution 
of $\Mgg$ at the right position expected from the edge in the scatter plot 
of $\Egam$ vs.\ $\gammas$, one can conclude that the observed events are 
really the signature of the two-body decay of $\neut$ into a photon and 
an invisible particle.

%
%
\section{Prospects for the neutralino decay search at VLHC}\label{sec:VLHC}
\begin{figure}[htb]
\begin{center}
\epsfbox[60 40 500 500]{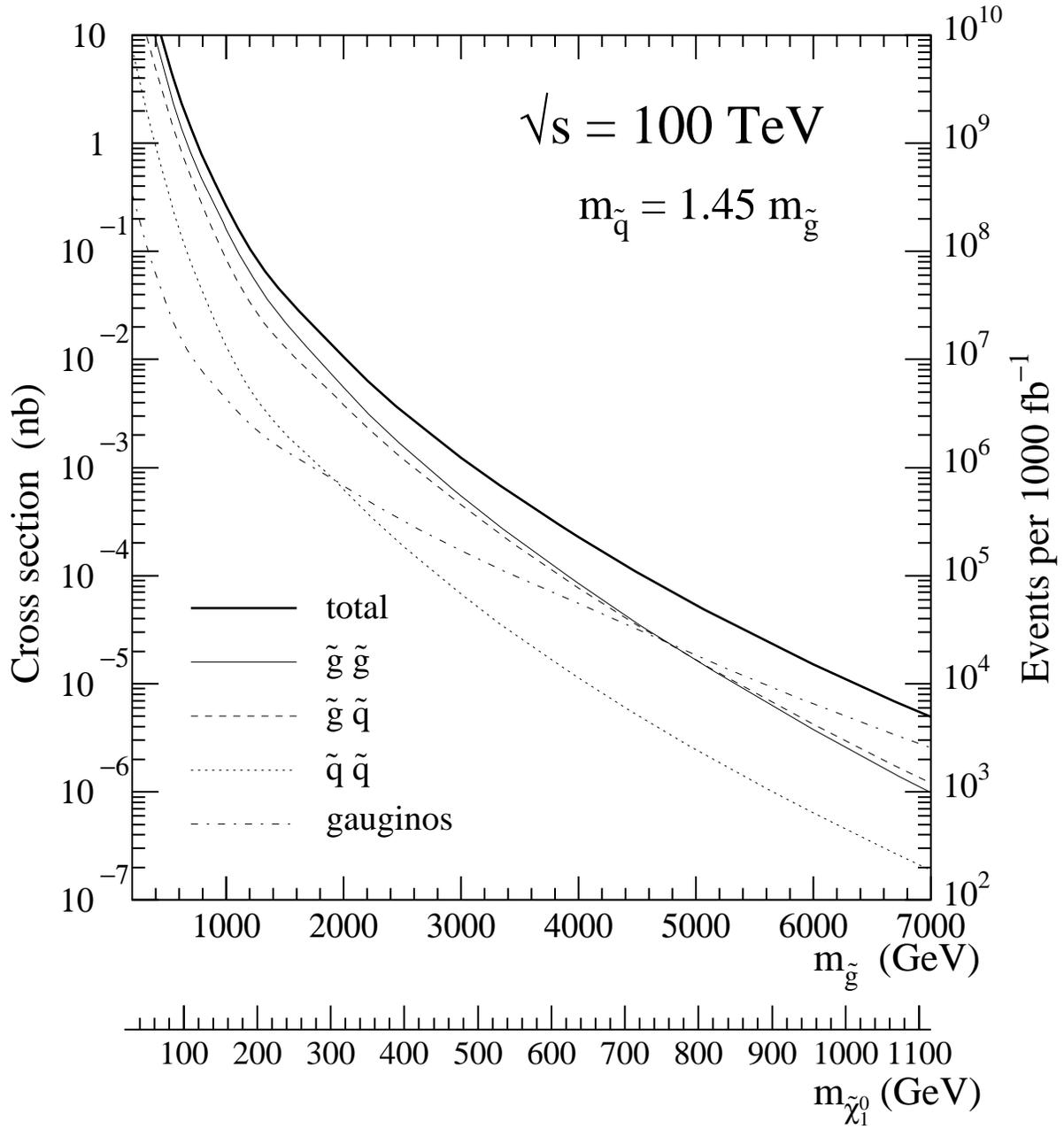}
\caption{
\protect\baselineskip 7mm
Total cross sections for the production of gluinos, squarks and gauginos 
in $pp$ collisions at $\rts=100$ TeV as a function of $\mglui$.  The 
corresponding values of $\mneut$ are also shown.  Also given in the 
vertical scale is the corresponding number of events per 1000 fb$^{-1}$.}  
\label{fig:vcross}
\end{center}
\end{figure}

Very Large Hadron Collider (VLHC) \cite{VLHC} has been discussed recently 
as a post-LHC machine, being operated at $\rts=100$--200 TeV with 
$\Lum=10^{34}$--$10^{35}$ cm$^{-2}$s$^{-1}$.  We stress here that 
the VLHC will become the neutralino factory much more powerful than the 
LHC\@.  Figure~\ref{fig:vcross} shows the total cross sections for the 
production of gluinos, squarks and gauginos, as well as the anticipated 
numbers of events per 1000 fb$^{-1}$, in $pp$ collisions at $\rts=100$ TeV 
as a function of $\mglui$.  Also shown in this figure are the corresponding 
values of $\mneut\approx\mglui/7$.  

\begin{figure}
\begin{center}
\epsfbox[60 40 500 500]{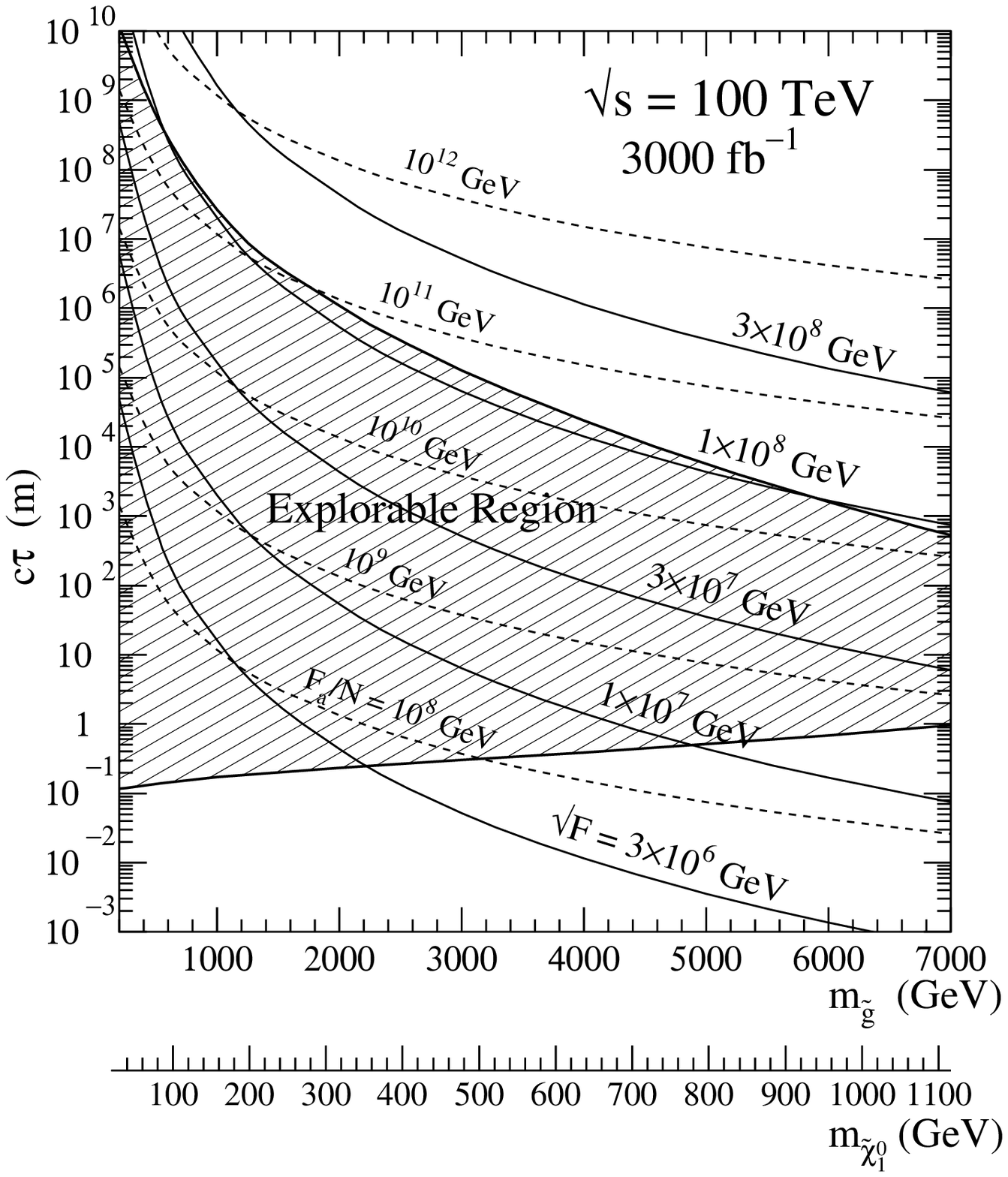}
\caption{
\protect\baselineskip 7mm 
Explorable range of $c\tau$ for the $\neut$ decay at VLHC with 3000 
fb$^{-1}$.  Also shown are the predicted curves of 
$c\tau(\neut\rightarrow\gamma\grav)$ for $\rtF=3\times10^{6}$ to 
$3\times10^{8}$ GeV (solid lines) and $c\tau(\neut\rightarrow\gamma\axi)$ 
for $\Fa/N=10^{8}$ to $10^{12}$ GeV (dashed lines).}
\label{fig:vreach}
\end{center}
\end{figure}

Figure~\ref{fig:vreach} shows the explorable range of $c\tau$ for the 
$\neut$ decay at the VLHC with 3000 fb$^{-1}$.  As in Fig.~\ref{fig:reach}, 
the upper end of the explorable region 
corresponds to the level of 10 decay events detected, and the lower one is 
derived from the condition that 100 decay events are detected.  
The explorable range of $c\tau$ is found to be 0.1 m to $1\times10^{7}$ 
km for $\mneut=25$ GeV, 0.2 m to $5\times10^{3}$ km for $\mneut=200$ GeV, 
and 1 m to 1 km for $\mneut=1000$ GeV\@.  We also show in this figure the 
predicted curves of $c\tau$ for the $\neut\rightarrow\gamma\grav$ and 
$\neut\rightarrow\gamma\axi$ decays for typical values of $\rtF$ and $\Fa$, 
respectively.  It is found that one can extend the reach for $\rtF$ up to 
$1\times10^{8}$ GeV through the $\neut\rightarrow\gamma\grav$ decay.  
In case of the $\neut\rightarrow\gamma\axi$ decay, $\Fa$ 
can be explored up to $\sim5\times10^{11}$ GeV for relatively light 
$\neut$.  It is worth noting that our experiment is unique in 
that one can explore such a high value of $\Fa$ in laboratories.  
In addition, our experiment can be performed under very high luminosity 
exceeding $10^{35}$ cm$^{-2}$s$^{-1}$, in contrast to usual collider 
experiments.

%
%
\section{Other possible searches}\label{sec:other}
\subsection{Other decay modes of the lightest neutralino}
One can search for other modes of the $\neut$ decay by using the setup 
proposed in this paper.  If the $\neut$ is sufficiently heavy, it can decay 
into $Z^{0}\grav$ or $h^{0}\grav$ (where $h^{0}$ is the Higgs boson) in the 
gravitino LSP scenario \cite{DDRT96,AKKMM96}.  In case the $\neut$ is 
higgsino-like, the $h^{0}\grav$ decay can even be dominant.  Since the 
$Z^{0}$ and $h^{0}$ 
bosons subsequently decay into $q\bar{q}$ or $l^{+}l^{-}$, the visible 
signature of these $\neut$ decays will be two hadron jets or two leptons 
without any signal in the anti-coincidence hodoscopes.  With the help of 
track detectors at the end of the decay tunnel, the decay point can be 
exactly located, which makes the kinematical analysis of the decay and the 
mass determination much more streightforward than in case of the 
$\neut\rightarrow\gamma\grav$ decay described in 
Section~\ref{sec:mass}.  The $\neut$ can also decay into such modes as 
$q\bar{q}^{\prime}l^{\pm}$, $q\bar{q}\nu$ and $\gamma\nu$ through weak 
violation of $R$-parity \cite{HS84}.  Furthermore, the decay of 
other longlived neutral particles, {\it e.g.}\ heavy neutrinos, can be 
detected with the same setup and method.  

\subsection{Slow decays of longlived charged particles}
Our setup can also be used to search for slow decays of heavy longlived 
charged particles.  
Although the existence of such particles would in principle be 
indicated in collider experiments by charged particle tracks with ionization 
larger than the minimum one, the signature of their decay will be missed if 
it occurs mostly outside the detector.  On the other hand, with our setup, 
one can clearly detect the decay signal of such longlived particles as in 
case of the $\neut$ decay.  The ``anti-coincidence'' hodoscopes 
can be used as the trigger counter.  

There are several candidates for longlived charged particles in SUSY 
models.  For example, if the gravitino or axino is the LSP, a charged 
superparticle such as the slepton ($\slep$) can be the 
NLSP, and may decay into $l^{-}\grav$ with $c\tau$ much longer than the 
detector size.  The chargino ($\chrg$) can be another example of the 
longlived NLSP if the mass difference between the $\chrg$ and the $\neut$ 
(assumed to be the LSP) is so small that it can decay only via 
$\chrg\rightarrow e^{-}\bar{\nu_{e}}\neut$ \cite{CDG97}.  Weak violation 
of $R$-parity would also allow the existence of 
longlived charged particles \cite{K96}.  In this case, the slepton can be 
the LSP, decaying slowly via, {\it e.g.}, $\slep\rightarrow l^{-}\nu$.  
Furthermore, heavy longlived charged particles could also exist in the 
SUSY breaking sector and messenger sectors of GMSB 
models \cite{AMR97,DGP96b}.  There are also many other speculations on 
heavy longlived charged particles 
[46--49].  

\subsection{Heavy stable charged particles}
If the $c\tau$ of heavy charged particles is extremely long, 
they will appear to be ``stable'' particles even in our experiment.  One 
can search for such particles with our setup by precisely measuring their 
time-of-flight over 40 m distance between the ``anti-coincidence'' 
hodoscopes and the shower detector at the end of the tunnel.  
The $\beta^{-1}$ distribution for heavy stable charged particles is expected 
to have a long tail following the peak at $\beta^{-1}\sim1$, in contrast to 
muons which make a sharp peak at $\beta^{-1}=1$.  Long flight path in our 
setup leads to excellent resolution of $\beta$, which will provide a clear 
signal for the heavy stable charged particle.  One could then determine 
its mass by measuring its momentum with magnetized iron and tracking 
chambers, which can be installed either in the active shield or at the very 
end of the shower detector.

%
%
\section{Conclusions}\label{sec:conc}
We have investigated the detection of a slow decay of the lightest 
neutralino (or any other longlived particles) at the CERN LHC and at VLHC\@.  
After pointing out that such hadron colliders will become the 
``neutralino factory'' producing $10^{6}$--$10^{9}$ $\neut$'s/yr if SUSY 
particles actually exist below $O(1)$ TeV, we have shown that a slow decay of 
$\neut$ can be detected in a dedicated experiment in which the collision 
products are dumped by a thick shield.  The decay product of $\neut$ can 
then be detected by a detector located at the end of a long decay tunnel.  
The slow arrival time and the large off angle $\psi$ of the decay product 
will provide a clear signature of slowly decaying $\neut$'s.  Considering 
the $\neut$ decay into a photon and a light invisible particle such as the 
gravitino ($\grav$) or axino ($\axi$), we find that one can explore the 
decay length ($c\tau$) of $\neut$ in a wide range, {\it i.e.}, 0.2 m to 
$1\times10^{5}$ km for $\mneut=25$ GeV and 1 m to 2 km for $\mneut=200$ GeV 
for the case of the LHC\@.  This corresponds to the range of the SUSY 
breaking scale $\rtF=2\times10^{5}$ to $2\times10^{7}$ GeV in case of the 
$\neut\rightarrow\gamma\grav$ decay.  At VLHC, one can extend the explorable 
range of $\mneut$ up to $\sim1000$ GeV, and that of $\rtF$ up to 
$\sim1\times10^{8}$ GeV\@.  In case of the $\neut\rightarrow\gamma\axi$ 
decay, one can explore the PQ symmetry breaking scale $\Fa$ up to 
$\sim5\times10^{11}$ GeV\@.  The mass of the decaying particle can be 
determined with an accuracy of $\lesssim20$\% using the correlation between 
the energy and the arrival time of the decay photon.  With the setup we 
propose, one can also search for (i)~other decay modes of $\neut$ such as 
$R$-parity violating one, (ii)~the decay of any other longlived neutral or 
charged particles, and (iii)~heavy stable charged particles.    

The experiment we proposed in this paper is a completely new type of 
collider experiment, which may turn out to be of crucial importance.  
The civil engineering of the intersection region at future hadron colliders 
has to be designed so that it can accommodate long decay tunnels for such 
experiments.

\section*{Acknowledgements}
All the calculations were performed on the RS/6000 workstations of the 
International Center for Elementary Particle Physics (ICEPP), University of 
Tokyo.  K.M. acknowledges a fellowship from the Japan Society for the 
Promotion of Science.  

\appendix
\section*{Appendixes}

%
%
\section{Realistic design for the shower detector}\label{sec:dete}
\begin{figure}
\begin{center}
\epsfbox[36 394 479 647]{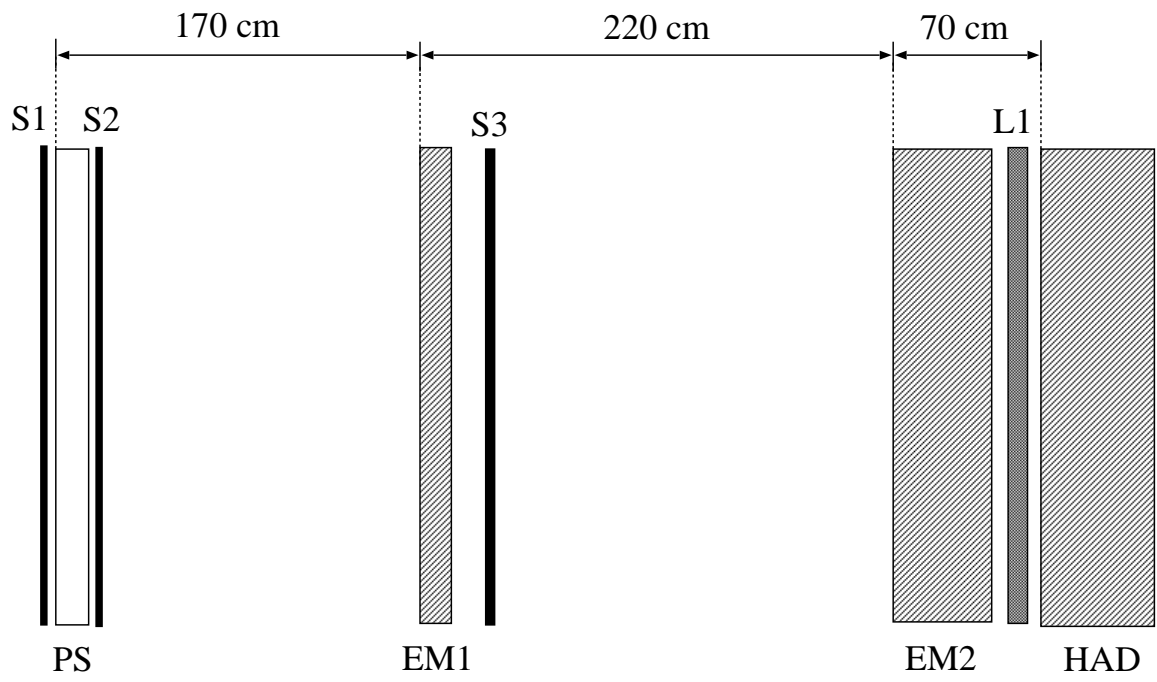}
\caption{
\protect\baselineskip 7mm
The schematic of the shower detector we investigated.  
It consists of the preshower detector 
(PS), two parts of the electromagnetic calorimeter (EM1 and EM2), three 
planes of plastic scintillators (S1, S2 and S3), the lead absorber (L1), 
and the hadron calorimeter (HAD).}
\label{fig:dete}
\end{center}
\end{figure}

Figure~\ref{fig:dete} shows a schematical view of a realistic design for the 
shower detector which we have investigated.  To separate 
electromagnetic showers initiated by photons from the 
background, we design the detector consisting of the preshower detector 
(PS), two parts of the electromagnetic calorimeter (EM1 and EM2), three 
planes of plastic scintillators (S1, S2 and S3), the lead absorber (L1), 
and the hadron calorimeter (HAD).  In addition, we can install the tracking 
chambers both in front of and behind the whole detector for the definite 
detection of high-energy muons (and possibly heavy stable charged 
particles).  
 
The preshower detector (PS) comprises two planes of multiwire proportional 
chambers (MWPCs) with cathode pad readout, with 2$\rl$ 
and 1$\rl$ lead converter in front, respectively (radiation length 
$\rl$ being 0.56 cm for lead).  The scintillator S1 in front of 
PS can work as the anti-coincidence counter for charged 
particles.  The scintillator S2, located just behind PS, is used 
for triggering.  The electromagnetic calorimeter is divided into two parts, 
each of which comprises modules with $10\times10$ cm$^{2}$ 
transverse dimensions.  The first part (EM1), which is located at 170 cm 
downstream of PS, consists of 5 layers of lead (1$\rl$) and scintillators 
(1 cm).  The second part (EM2), 
which is located at 220 cm downstream of EM1, consists of 20 layers of lead 
(1$\rl$) and scintillators (1 cm).  Total radiation lengths of EM1 and EM2 
are 5.1$\rl$ and 20.4$\rl$, respectively.  The thick scintillator S3, 
located between EM1 and EM2, is used not only for triggering but also for 
the measurement of the arrival time.  
The hadron calorimeter (HAD) contains 15 layers of 5 cm thick iron and 2 cm 
thick scintillators.  Total thickness is $\sim5$ nuclear interaction lengths.  
Hadronic showers will 
deposit much of its energy in HAD\@.  Between EM2 and HAD, the lead absorber 
L1 of 20$\rl$ in thickness is inserted to totally absorb electromagnetic 
showers.  

A coincidence of two planes of scintillators, {\it i.e.}\ S2 and S3, 
constitutes the first trigger.  We also require the signals 
from EM1 and EM2 to exceed a threshold, which leads to the rejection of 
(i)~muons going through the detector as minimum ionizing particles, and 
(ii)~hadronic showers initiated by high-energy neutrinos in HAD\@.  

\begin{figure}
\begin{center}
\epsfbox[-10 -30 425 425]{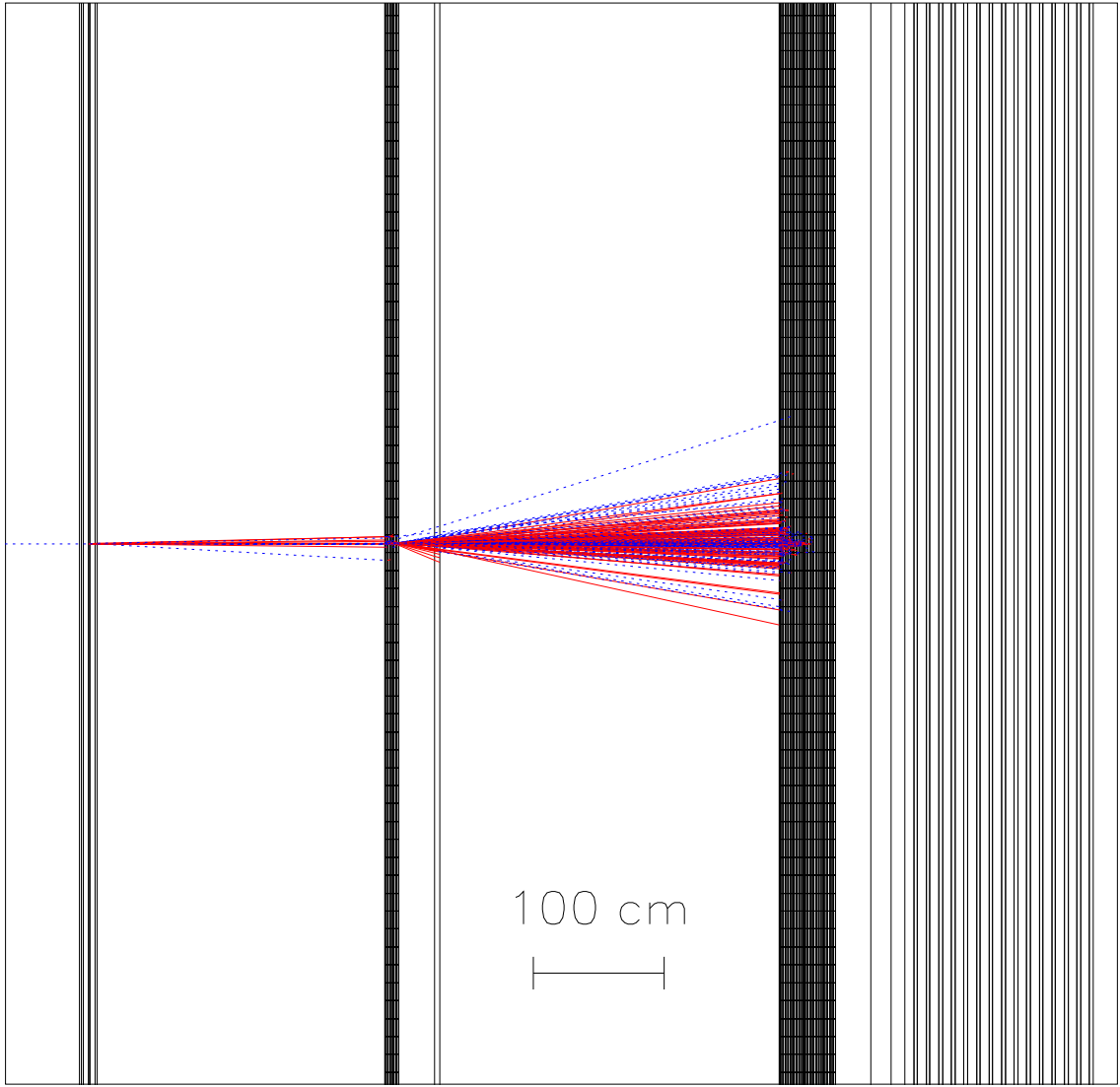}
\caption{
\protect\baselineskip 7mm
An example of the simulated electromagnetic shower which is initiated by 
100 GeV photon incident on the shower detector.  Solid and dashed lines 
indicate tracks of $\geq50$ MeV electrons and photons, respectively.}
\label{fig:event}
\end{center}
\end{figure}

\begin{figure}
\begin{center}
\epsfbox[30 30 480 480]{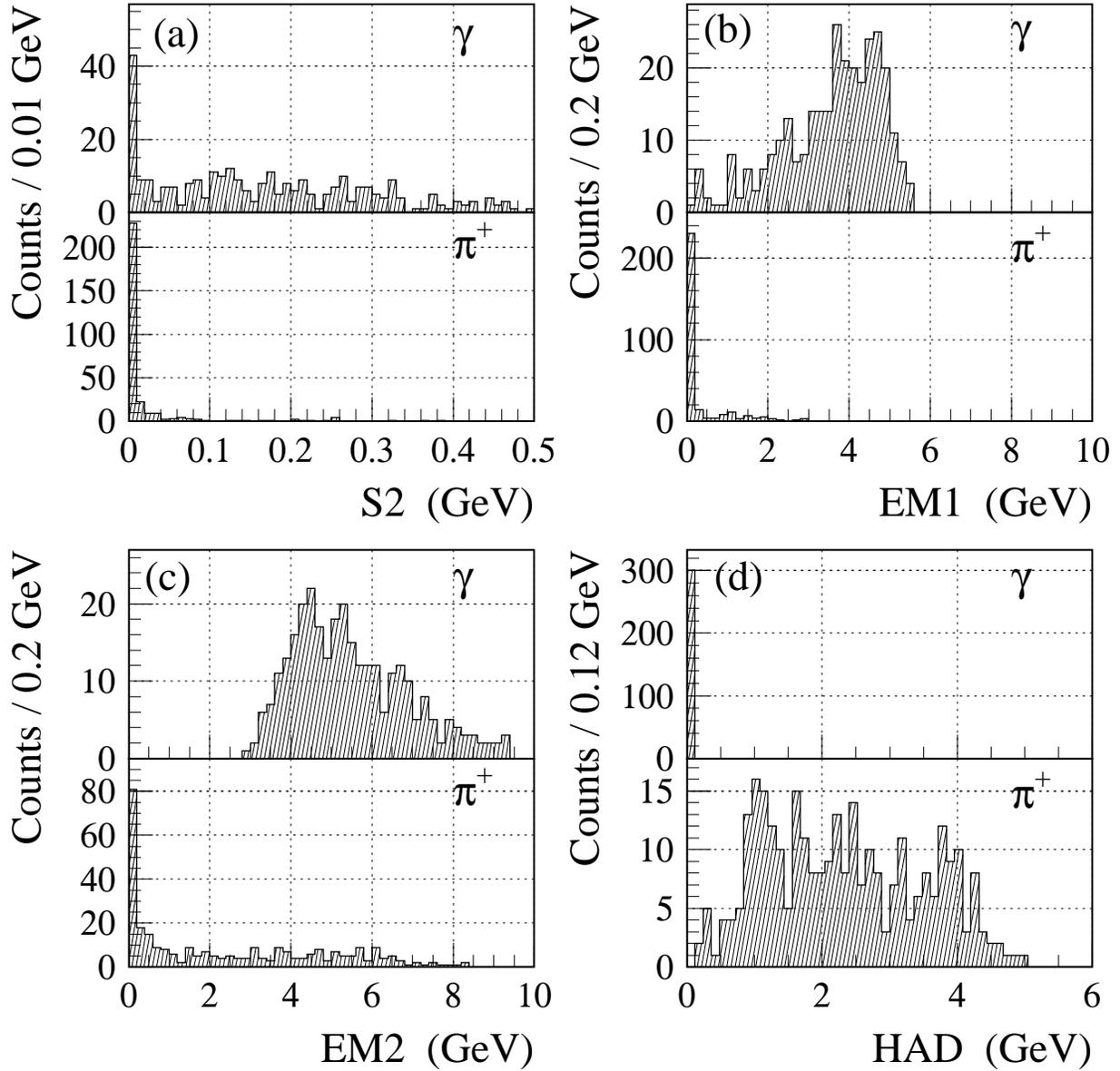}
\caption{
\protect\baselineskip 7mm
Distributions of energy deposited by 100 GeV photons and 
charged pions in (a)~S2, (b)~EM1, (c)~EM2, and (d)~HAD.}
\label{fig:pid}
\end{center}
\end{figure}

For this detector configuration, we have performed Monte Carlo simulations 
using GEANT 3.21 \cite{GEANT}.  A typical event for 100 GeV photon incident 
on the shower detector is shown in Fig.~\ref{fig:event}.  We can see that 
the electromagnetic shower starts to evolve in PS, reaches its maximum after 
passing EM1, and is mostly absorbed in EM2.  Figure~\ref{fig:pid} shows the 
distributions of energy deposited in active layers of (a)~S2, (b)~EM1, 
(c)~EM2, and (d)~HAD, by 100 GeV photons and charged pions which are 
incident at $0^{\circ}$ on the shower detector.  It can be seen that 
electromagnetic showers (initiated by photons) can clearly be separated from 
hadronic showers (initiated by pions) using the deposit energy of each 
component.  

\begin{figure}
\begin{center}
\epsfbox[40 40 500 530]{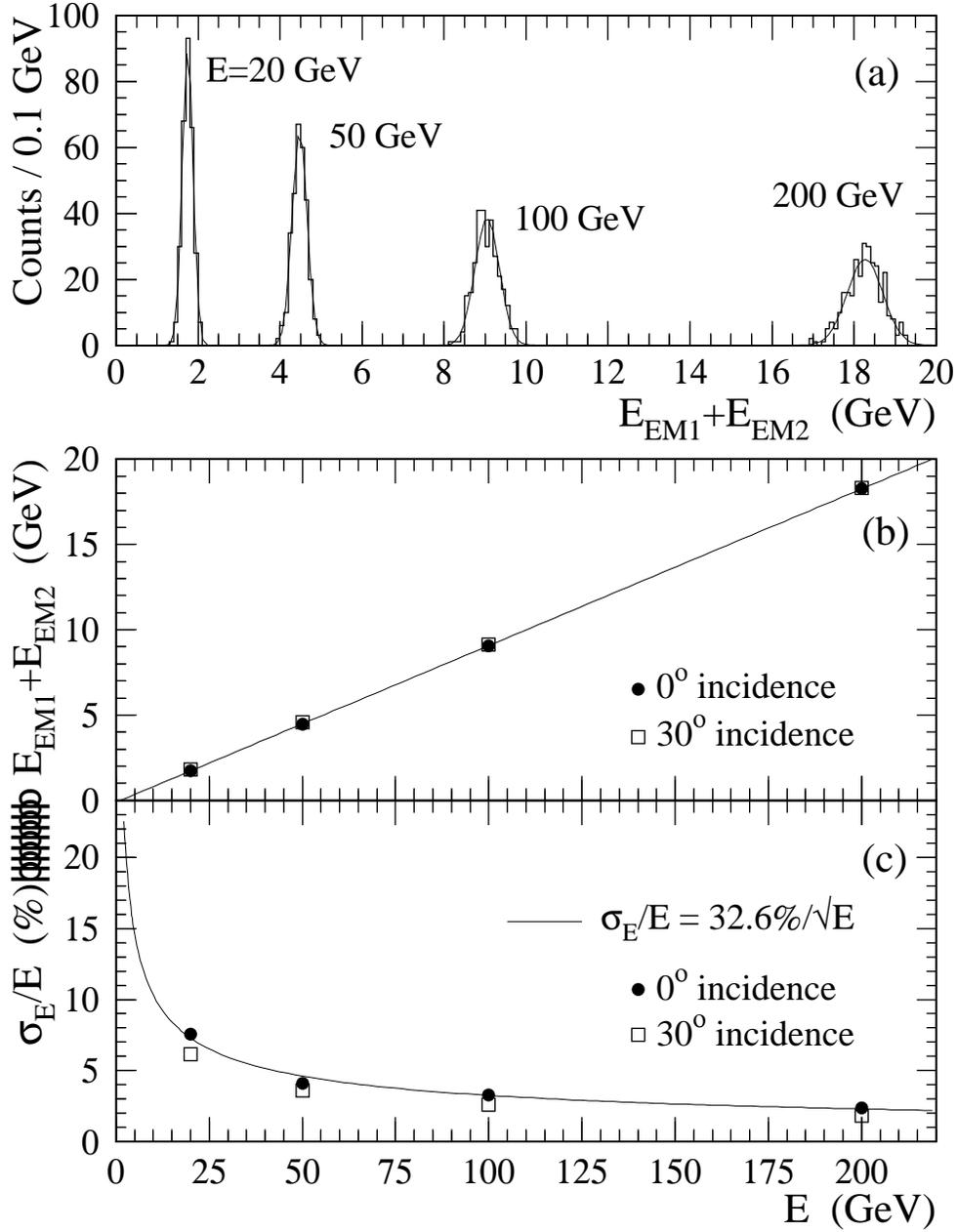}
\caption{
\protect\baselineskip 7mm
(a)~Distributions of the sum of deposit energy in EM1 and EM2 for photons 
incident at $0^{\circ}$ on the shower detector with energy $E=20$, 50, 100 
and 200 GeV\@.  The results of the fit by a gaussian are also shown.  
(b)~The sum of deposit energy in EM1 and EM2 vs.\ incident photon 
energy.  
(c)~The energy resolution as a function of incident photon energy.  
The curve is the result of the fit for $0^{\circ}$ data.}
\label{fig:eneres}
\end{center}
\end{figure}

The energy of incident photons can be determined from the sum of deposit 
energy in active layers of EM1 and EM2, denoted as $E_{\rm EM1}+E_{\rm EM2}$, 
whose distributions are shown in Fig.~\ref{fig:eneres}(a) for photons 
incident at $0^{\circ}$ with energy $E=20$, 50, 100 and 200 GeV\@.  The 
inefficiency concerning the scintillation light collection is not included.  
Figure~\ref{fig:eneres}(b) shows the relation between $E$ and 
$E_{\rm EM1}+E_{\rm EM2}$ for $0^{\circ}$ and $30^{\circ}$ incidence.  It can 
be seen that good linearity is realized for $E=20$--200 GeV\@.  The resultant 
energy resolution of the shower detector as a function of incident photon 
energy is shown in Fig.~\ref{fig:eneres}(c).  The fit result for $0^{\circ}$ 
data is $\sigma_{E}/E=32.6\%/\sqrt{E \;({\rm GeV})}$ for $E=20$--200 GeV\@.  

\begin{figure}
\begin{center}
\epsfbox[60 30 500 500]{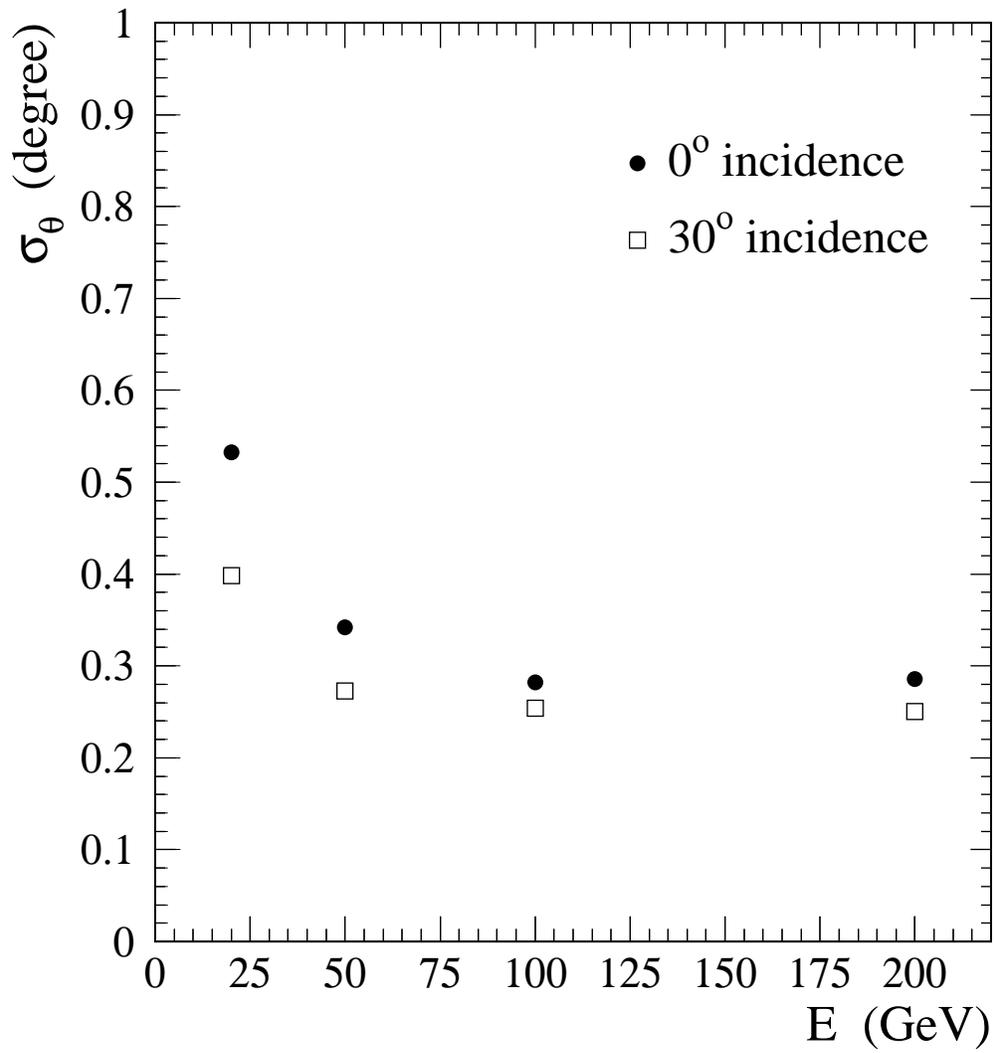}
\caption{
\protect\baselineskip 7mm
The angular resolution of the reconstructed shower direction as a function of 
incident photon energy.}
\label{fig:angres}
\end{center}
\end{figure}

The shower direction is obtained from the reconstructed position of the shower 
in PS and that in EM2.  The position of the shower in PS can be determined 
from the pad signal of MWPCs, whose spatial resolution is assumed to be 
5 mm.  The shower center in EM2 can be estimated from the observed division 
of energy within a local cluster of modules.  We find that good spatial 
resolution ($\sigma_{x,y}\sim2$ cm) in comparison with the size of the 
module ($10\times10$ cm$^{2}$) can be obtained owing to the large gap between 
EM1 and EM2, which makes the shower spread widely before entering EM2.  The 
resultant angular resolution averaged over various incident positions is 
shown in Fig.~\ref{fig:angres}.  It is found to be 
$\sigma_{\theta}\lesssim0.3^{\circ}$ for photons with $E=50$--200 GeV\@.  
The angular resolution could be improved if tracking chambers are used 
as active layers of EM1 and EM2 instead of scintillators.

%
%
\section{Rates of high-energy neutrinos and muons}\label{sec:munu} 
Although a large number of particles will be produced at the LHC, most of 
them (hadrons, $e^{\pm}$ and $\gamma$) will be dumped in our setup by the 
iron shield with thickness of $\sim60$ nuclear interaction lengths.  However, 
high-energy neutrinos and muons can go through the shield and might interact 
with the material of the detector.  The main source of high-energy neutrinos 
and muons is expected to be ``prompt'', {\it i.e.}, come from semileptonic 
decays of bottom and charm hadrons \cite{ATLAS}, while the contribution from 
the decay of $\pi$ and $K$ mesons is negligible at least in our setup.  In 
order to estimate the rates of these prompt neutrinos and muons, we have 
performed Monte Carlo simulations using ISAJET 7.13 \cite{ISAJET} for $pp$ 
collisions at $\rts=14$ TeV, by taking the total cross section of the bottom 
plus charm production to be 2 mb.  Total $2.4\times10^{6}$ events are 
generated, and the neutrinos and muons which point to the detector covering 
a polar angle of $\thtun\pm10^{\circ}$ and an elevation angle of 
$\pm10^{\circ}$ are recorded.  We consider four cases, {\it i.e.}, 
$\thtun=15^{\circ}$, $20^{\circ}$, $25^{\circ}$ and $30^{\circ}$.  

\subsection{Interacting neutrinos}
\begin{figure}
\begin{center}
\epsfbox[60 30 500 500]{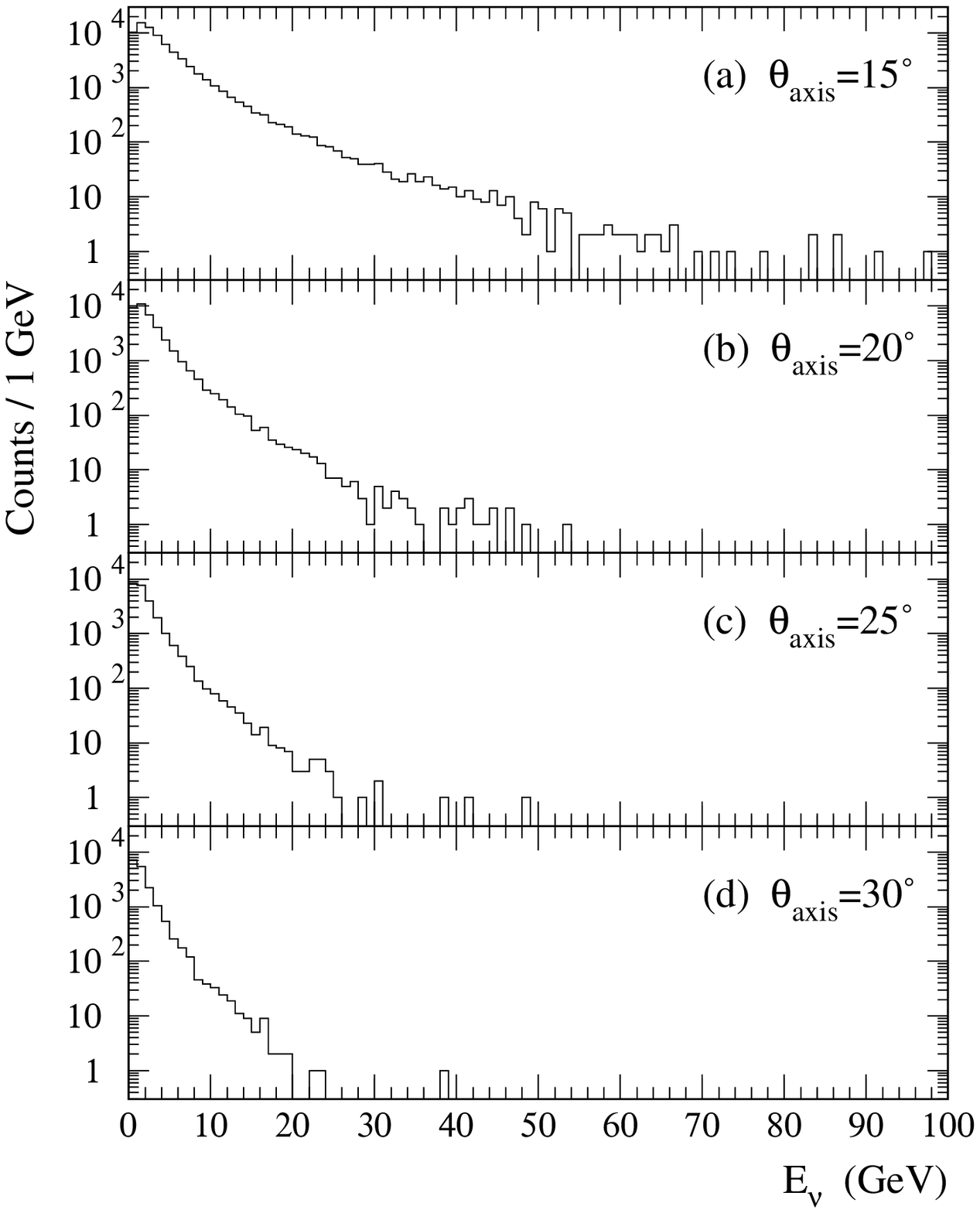}
\caption{
\protect\baselineskip 7mm
The energy spectra of the prompt neutrinos produced in the decay of bottom 
and charm hadrons for total $2.4\times10^{6}$ generated events.  The 
polar angle of the tunnel axis $\thtun$ is taken to be (a)~$15^{\circ}$, 
(b)~$20^{\circ}$, (c)~$25^{\circ}$, and (d)~$30^{\circ}$.}
\label{fig:nu}
\end{center}
\end{figure}

The energy spectra of the prompt neutrinos are shown in Fig.~\ref{fig:nu}.  
The spectra have a long high-energy tail when $\thtun$ is small.  The 
neutrino could mimic the $\gamma\grav$ or $\gamma\axi$ signal if it interacts 
in the preshower detector and deposits enough energy (see 
Appendix~\ref{sec:dete} for the detailed design for the shower detector).  
As a simple estimate of such background, we calculate the number of 
neutrinos which have energy $\Enu\geq20$ GeV and interact with 
$3\rl\approx19$ g/cm$^{2}$ lead.  The interaction cross section is taken 
to be $5\times10^{-39}(\Enu/{\rm GeV})$ cm$^{2}$/nucleon.  The estimated 
numbers are $\sim0.2$, $\sim0.02$, $\lesssim0.01$ and $\lesssim0.001$ per 
100 fb$^{-1}$ for $\thtun=15^{\circ}$, $20^{\circ}$, $25^{\circ}$ and 
$30^{\circ}$, respectively.  Thus the background induced by the prompt 
neutrinos is negligibly small.  

\subsection{Muon-induced background}
\begin{figure}
\begin{center}
\epsfbox[60 30 500 500]{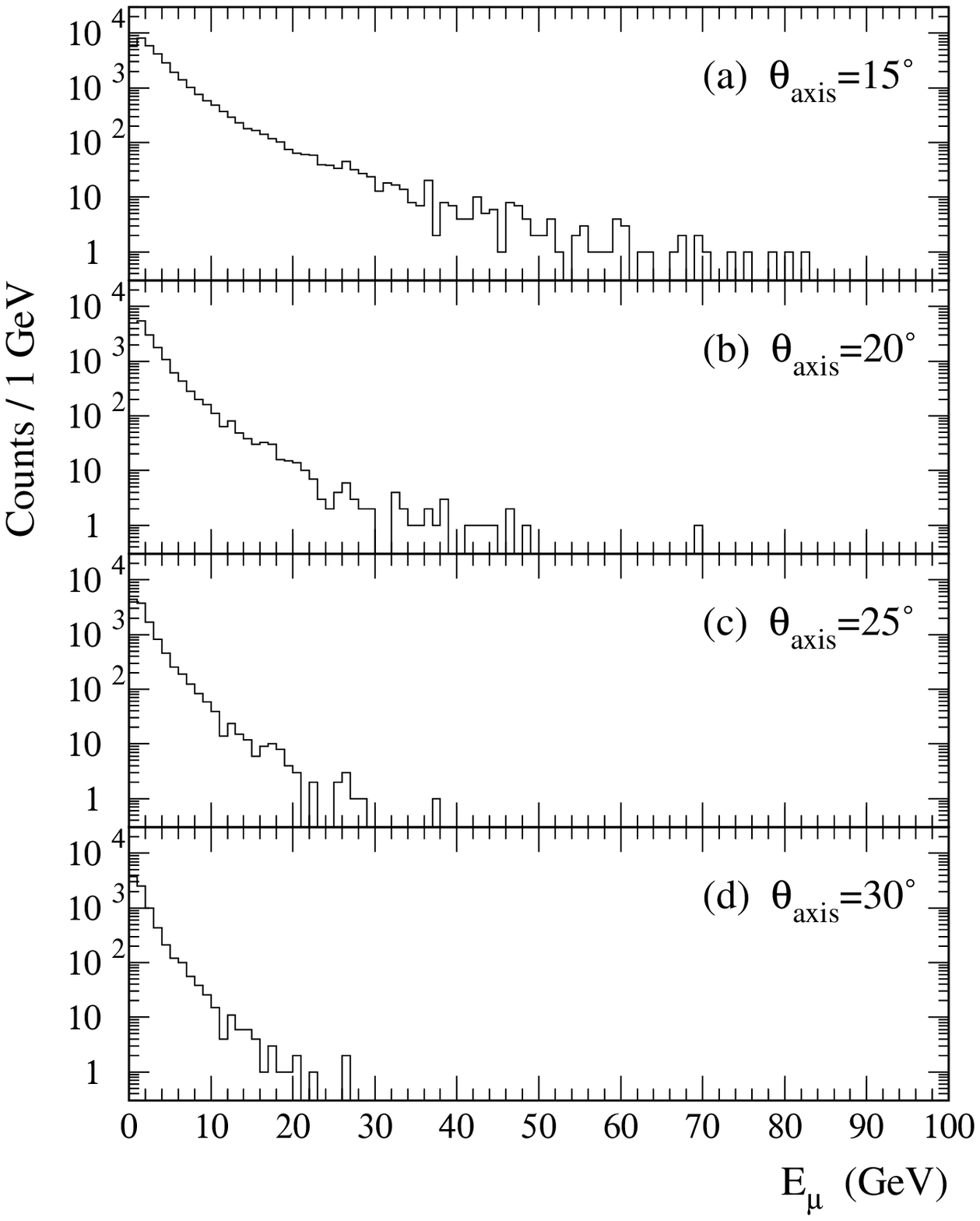}
\caption{
\protect\baselineskip 7mm
The energy spectra of the prompt muons produced in the decay of bottom 
and charm hadrons for total $2.4\times10^{6}$ generated events.  The 
polar angle of the tunnel axis $\thtun$ is taken to be (a)~$15^{\circ}$, 
(b)~$20^{\circ}$, (c)~$25^{\circ}$, and (d)~$30^{\circ}$.}
\label{fig:mu}
\end{center}
\end{figure}

Figure~\ref{fig:mu} shows the energy spectra of the prompt muons.  Only the 
muons with $\Emu\geq20$ GeV can go through the shield because of the 
ionization energy loss.  The resultant rates of muons entering the tunnel 
are then estimated to be $\sim5$ kHz, $\sim600$ Hz, $\lesssim200$ Hz and 
$\lesssim80$ Hz for $\thtun=15^{\circ}$, $20^{\circ}$, $25^{\circ}$ and 
$30^{\circ}$, respectively, with luminosity of $10^{34}$ cm$^{-2}$s$^{-1}$.  
The single rate of the prompt muons will thus be comfortably small if we 
take $\thtun\gtrsim25^{\circ}$. 
   
High-energy muons can deposit a significant portion of their 
energy by bremsstrahlung or direct $\ee$ pair production, both of which 
could lead to the electromagnetic shower.  When combined with inefficiencies 
in the anti-coincidence counters, these muons might in principle mimic the 
$\gamma\grav$ or $\gamma\axi$ signal.  Folding the simulated spectra by the 
showering probability \cite{T74} of muons, we estimate the numbers of 
muon-induced high-energy ($E\geq10$ GeV) showers starting to evolve in the 
preshower detector to be $\sim10^{7}$, $\sim10^{6}$, $\lesssim2\times10^{5}$ 
and $\lesssim10^{5}$ per 100 fb$^{-1}$ for $\thtun=15^{\circ}$, $20^{\circ}$, 
$25^{\circ}$ and $30^{\circ}$, respectively.  Assuming a veto inefficiency of 
$\sim10^{-6}$ by the staggered layers of anti-coincidence scintillators, the 
muon-induced background is expected to be negligibly small for 
$\thtun\gtrsim25^{\circ}$.

Even if the rate of such muon-induced background turns out to be 
non-negligible, off-line analysis will eliminate such background based on 
the information on the arrival time as well as on the off angle $\psi$.  

\newpage
%
%

\begin{figure}
\begin{center}
\epsfbox[60 30 500 500]{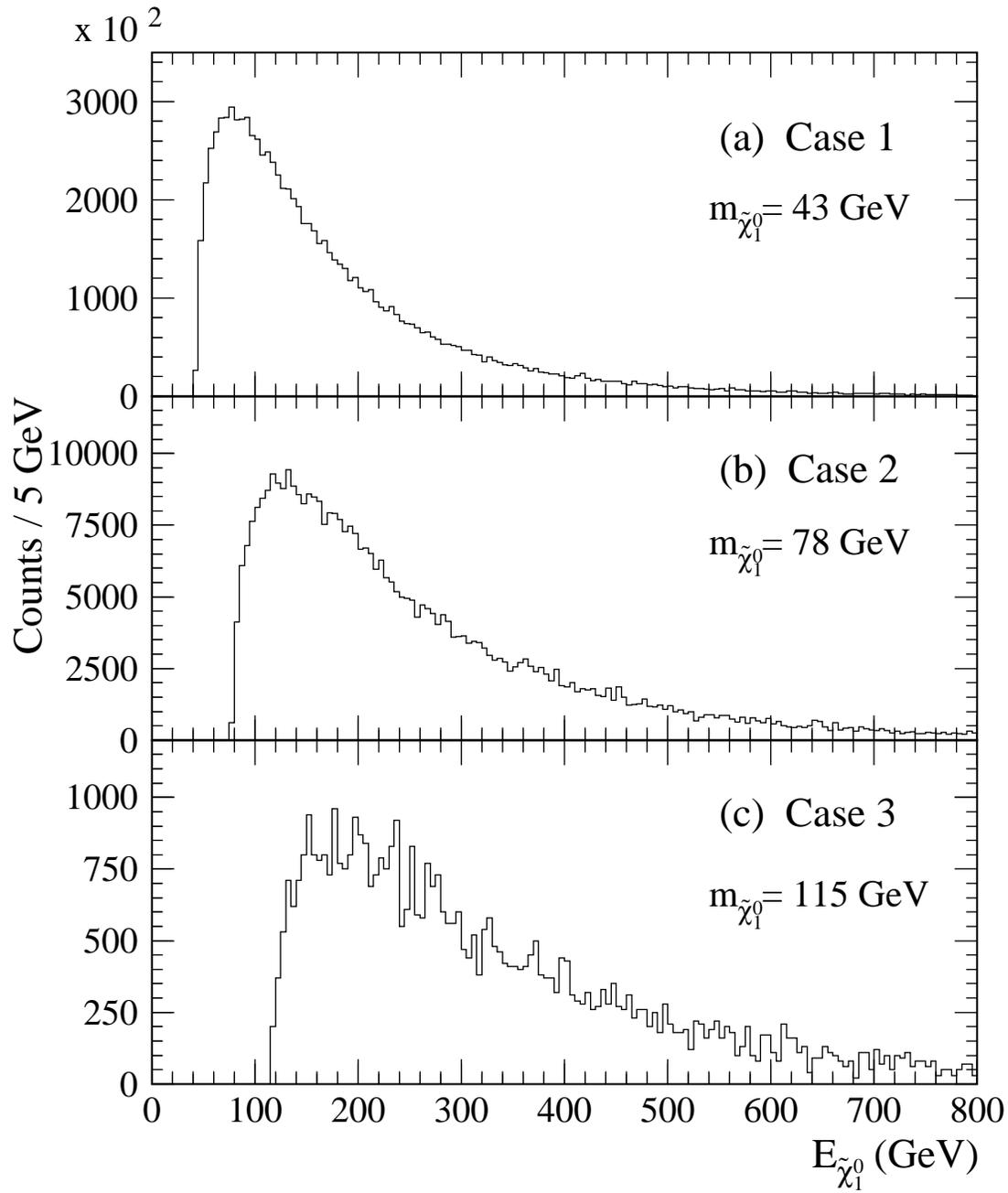}
\caption{
\protect\baselineskip 7mm
The energy spectra of $\neut$'s which point to the detector for 
(a)~Case~1, (b)~Case~2, and (c)~Case~3.}
\label{fig:eneut}
\end{center}
\end{figure}

\vspace*{3cm}

\clearpage

\end{document}